\documentclass[11pt,sort&compress]{elsarticle}
\usepackage[paper=a4paper,marginratio={1:1,5:6}]{geometry}
\makeatletter
\def\ps@pprintTitle{%
 \let\@oddhead\@empty
 \let\@evenhead\@empty
 \def\@oddfoot{\centerline{\thepage}}%
 \let\@evenfoot\@oddfoot}
\makeatother
\usepackage{natbib}
\usepackage[hyperref]{xcolor}
\definecolor{darkgreen}{rgb}{0.01, 0.75, 0.24}
\definecolor{darkblue}{HTML}{2B66D3}
\usepackage[colorlinks=true,
            linkcolor=darkblue,
            urlcolor=darkblue,
            citecolor=darkblue,linkcolor=darkblue]{hyperref}

\usepackage{amsmath}
\usepackage{amsfonts}
\usepackage{siunitx}
\usepackage{slashed}
\usepackage{accents}
\usepackage{amssymb}
\usepackage{mathrsfs}
\usepackage{mathtools}
\usepackage[utf8]{inputenc}
\usepackage[T1]{fontenc}
\let\oldbibliography\thebibliography
\renewcommand{\thebibliography}[1]{%
  \oldbibliography{#1}%
  \setlength{\itemsep}{1.4pt}%
}

\usepackage[cal=boondox,calscaled=1]{mathalfa}
\DeclareMathAlphabet{\bbvar}{U}{BOONDOX-ds}{m}{n}
\DeclareMathAlphabet{\mathsl}{\encodingdefault}{\familydefault}{m}{sl}

\usepackage{psfrag}
\usepackage{pstool}
\usepackage{caption}
\usepackage{tabularx}
\usepackage{multirow}
\usepackage{pbox}
\usepackage{graphicx}
\usepackage{relsize}
\newcommand*{\smallcirc}{\text{$\mathrel{\mathsmaller{\mathsmaller{\circ}}}$}}

\newcommand{\qq}[1]{``#1''} 
\newcommand{\utilde}[1]{\underaccent{\tilde}{#1}}
\newcommand{\di}{\mathrm{d}}
\usepackage{tensor}
\newcommand{\ou}[3]{\tensor{#1}{^{#2}_{#3}}}
\newcommand{\uo}[3]{\tensor{#1}{_{#2}^{#3}}}

\newcommand{\I}{\mathrm{i}} 
\newcommand{\E}{\mathrm{e}} 
\newcommand{\CC}{\mathrm{cc.}} 
\newcommand{\C}{\mathbb{C}}
\newcommand{\N}{\mathbb{N}}
\newcommand{\R}{\mathbb{R}}
\newcommand{\Z}{\mathbb{Z}}

\newcommand{\1}{\mathnormal{1}}
\newcommand{\0}{o}

\newcommand{\eref}[1]{(\ref{#1})}

\newcommand{\doi}[1]{\href{https://doi.org/#1}{{\tt doi:#1}}}

\DeclareMathAlphabet{\bbgreek}{U}{bbold}{m}{n}

\newcommand{\mtext}[1]{\text{\it #1}}
\usepackage{tikz-cd}
\usetikzlibrary{arrows.meta, automata, 
                positioning, 
                quotes}
\usepackage[low-sup]{subdepth}

\newcommand\vpm{\mathbin{\vcenter{\hbox{
  \oalign{\hfil$\scriptstyle+$\hfil\cr
          \noalign{\kern-.3ex}
          $\scriptscriptstyle({-})$\cr}}}}}
\DeclareMathAlphabet{\sfit}{OT1}{fos}{sb}{it}
\DeclareMathAlphabet{\mathsf}{OT1}{fos}{sb}{n}

\definecolor{darkgreen}{rgb}{0.01, 0.75, 0.24}
\definecolor{darkblue}{HTML}{2B66D3}

\newcommand{\normord}[1]{{\boldsymbol{:}\mathrel{#1}\boldsymbol{:}}}
\usepackage[multiple, flushmargin]{footmisc}
\let\originalleft\left
\let\originalright\right
\renewcommand{\left}{\mathopen{}\mathclose\bgroup\originalleft}
\renewcommand{\right}{\aftergroup\egroup\originalright}

\usepackage{scalerel}

\newcommand{\dbarvar}{{\mathrm{d}\mkern-7.5mu\lower.18ex\hbox{$\textasciitilde$}\mkern-1.5mu}}

\newcommand{\Tr}{\operatorname{Tr}}

\renewcommand{\emph}[1]{{\it #1}}

\begin{document}

\begin{abstract}
\noindent Starting from the symplectic potential for the $\gamma$-Palatini--Holst action on a null hypersurface, we identify an auxiliary conformal field theory (CFT), which carries a representation of the constraint algebra of general relativity on a null surface. The radiative data, which is encoded into the shear of each null generator, is mapped into an $SU(1,1)$ current algebra on each light ray. We study the resulting quantum theory for both bosonic and fermionic representations. In the fermionic representation, the central charge on each null ray is positive, for bosons it is negative. 
A negative central charge implies a non-unitary CFT, which has negative norm states. In the model, there is a  natural $SU(1,1)$ Casimir. For the bosonic representations, the $SU(1,1)$ Casimir can have either sign. For the fermionic representations, the $SU(1,1)$ Casimir is always greater or equal to zero. To exclude negative norm states, we restrict ourselves to the fermionic case. To understand the physical implications of this restriction, we express the $SU(1,1)$ Casimir in terms of the geometric data. In this way, the positivity bound on the $SU(1,1)$ Casimir translates into an upper bound for the shear of each null generator. In the model, this bound must be satisfied for all three-dimensional null hypersurfaces. This in turn suggests to apply it to an entire null foliation in an asymptotically flat spacetime. In this way, we obtain a bound on the radiated power of gravitational waves in the model. 
%
 %
  \end{abstract}%
\title{Quantum Geometry of the Light Cone: Fock representation and Spectrum of Radiated Power}
\author{Wolfgang Wieland}
\address{Institute for Quantum Gravity, Theoretical Physics III, Department of Physics\\Friedrich-Alexander-Universität Erlangen-Nürnberg\\Staudstraße 7, 91052 Erlangen. Germany
\\{\vspace{0.5em}\normalfont \today}}

\maketitle
\vspace{-1.5em}
\hypersetup{
  linkcolor=black,
  urlcolor=black,
  citecolor=black
}
{\tableofcontents}\vspace{-0.5em}
\hypersetup{
  linkcolor=darkblue,
  urlcolor=darkblue,
  citecolor=darkblue,
}
\begin{center}{\noindent\rule{\linewidth}{0.4pt}}\end{center}\newpage
\section{Introduction}
\noindent In this research, we improve on earlier results on our recent proposal \cite{Wieland:2024dop,Wieland:2025LP} for a non-perturbative and quasi-local quantisation of gravitational null initial data. Compared to the earlier model, we present two main developments along smaller technical details. Originally, the model was motivated by the observation that classical general relativity admits exact solutions, in which the metric is distributional \cite{Aichelburg1971,Balasin:2007gh,Luk:2012hi,Griffiths:1991zp,GRColombBook}. In \cite{Wieland:2024dop,Wieland:2025LP}, we considered a truncation of  classical null initial data to piecewise constant shear. We quantized the resulting reduced phase space and identified the physical states of the model.
In the following, we drop this restriction. Utilizing a standard Fock representation, we extend the model to arbitrary profiles of the free radiative data. Thus, each light ray carries infinitely many  modes. The second improvement concerns the spectrum of the radiated power. In \cite{Wieland:2025LP}, we identified a critical luminosity
\begin{equation}
\mathcal{L}_{\mtext{crit.}}=\frac{c^5}{G}\frac{1}{1+\gamma^2}\leq\frac{c^5}{G},
\end{equation}
where $\gamma$ is the Barbero--Immirzi parameter \cite{Immirziparam,Barbero1994}, which determines the fundamental \emph{area gap} of loop quantum gravity \cite{Ashtekar:2021kfp,status, thiemann, rovelli}. In the model \cite{Wieland:2025LP}, the critical luminosity splits the spectrum of the radiated power into two parts. For \emph{Infra-Planckian states} that lie below the critical threshold, the spectrum is discrete. \emph{Ultra-Planckian states}, which are states that lie above the critical threshold, have a continuous spectrum. We then argued that such states suffer from a pathology that does not appear below the critical luminosity \cite{Wieland:2025LP}. 
In the following, we employ conformal field theory methods to strengthen this conjecture.\smallskip

Our results build very heavily on a number of developments in the area, some of which are very recent. First of all, the quantisation utilizes earlier methods on isolated horizons \cite{Ashtekar:2000eq,isohorizon,Ashtekar:1999wa,Ashtekar:aa,BarberoG.:2012ae,DiazPoloPranzetti,Ashtekar:2001is,Engle:2010kt} and the bosonic spinor representation of spin network states \cite{Girelli:2005ii,Bianchi:2016hmk,Borja:2010rc,twist,twistconslor,komplexspinors}. Recent research on \emph{edge modes}, \emph{boundary symmetries} and local subsystems in gauge theories and gravity \cite{Balachandran:1995qa,PhysRevD.51.632,Donnelly:2016auv,Gomes:2016mwl,Geiller:2017xad,Speranza:2017gxd,Geiller:2017whh,Wieland:2017cmf,Wieland:2017zkf,Donnelly:2016rvo,Takayanagi:2019tvn,Freidel:2019ees,Francois:2021aa,Freidel:2020ayo,Freidel:2020svx,Freidel:2020xyx,Donnelly:2020xgu,Freidel:2021cjp,Carrozza:2021gju,Carrozza:2022xut,Kabel23,Giesel:2024xtb,Ciambelli:2022vot,Wieland:2021vef,Araujo-Regado:2024dpr} plays an important in this research as well. Most crucially, the model expands on developments on field theories on null hypersurfaces \cite{Donnay:2019jiz,Ciambelli:2019lap,Ciambelli:2018ojf,Lehner:2016vdi,Parattu:2015gga}, symplectic geometry of gravitational null initial data \cite{Hopfmuller:2016scf,Freidel:2022vjq,Adami:2021nnf,Ciambelli:2023mir,Reisenberger:2018xkn,Fuchs:2017jyk,Reisenberger:2012zq,AndradeeSilva:2022iic} and the role of conformal field theories in its quantum description \cite{Ciambelli:2024swv,Strominger:1997eq,Carlip:1998wz,Carlip:1999cy}. 

\paragraph{Outline} Our starting point (\hyperref[sec2]{Section 2}) is the classical phase space of the $\gamma$-Palatini--Holst action in a compact domain with a boundary that is null. This phase space describes a local gravitational subsystem \cite{Donnelly:2016auv}. Blowing this subsystem up into an infinite volume, we can reach the asymptotic radiative phase space \cite{Wieland:2020gno}. The quantisation of the asymptotic phase space returns the usual $S$-matrix theory of perturbative quantum gravity \cite{Ashtekar:1981sf,Strominger:2017zoo}. In \hyperref[sec3]{Section 3}, we consider the quantisation of the subsystem at finite distance using conformal field theory (CFT) methods. Each null ray carries an auxiliary CFT with a central charge. The value of the central charge depends on the statistics of the auxiliary field theory from which we construct a representation of an $SU(1,1)$ current algebra on the lightcone. To guarantee unitary of the inner product, we are led to choose a fermionic representation in which the $SU(1,1)$ currents are built from the tensor product of fermionic modes, which are charged under $SU(1,1)$. By restricting ourselves to the fermionic case, the central charge is positive. \hyperref[sec4]{Section 4} explains how the spectrum of the radiated power depends on the sign of the central charge. If the central charge is positive, the radiated power is bounded from above. This is in conflict with the standard perturbative quantisation of gravitational null initial data at future (past) null infinity. On the asymptotic Fock space of the perturbative $S$-matrix approach, there is no bound on the radiated power. The results of the paper are an implicit proof that the semi-classical $\hbar\rightarrow 0$ and the asymptotic $r\rightarrow \infty$ limit may not commute.  This could have important consequences. The Planck power is independent of $\hbar$. It is, therefore, a highly unexpected result that non-perturbative effects of  quantum gravity can, in principle, create a bound on the radiated power. In short, quantum effects can appear at scales not considered relevant for quantum gravity before.

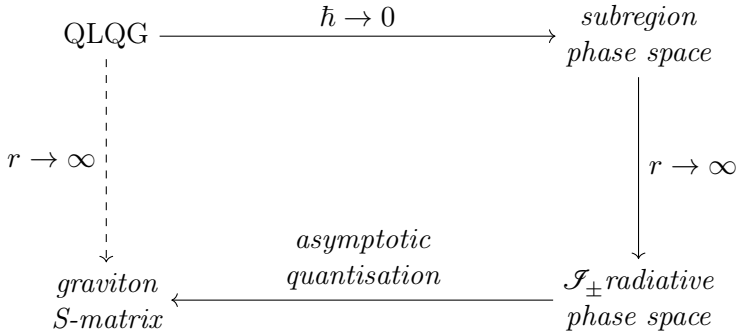
\begin{figure}[h]
\begin{tikzpicture}[node distance = 35mm and 70mm,
 auto,every edge quote/.style = {auto, font=\small}]
  \node (QG) {QLQG};
  \node (X) [right of=QG,align=center]  {};
  \node (LG) [right of=X,align=center] {\emph{subregion}\\\emph{phase space}};
  \node (SMT) [below of=QG,align=center] {\emph{graviton}\\\emph{S-matrix}};
   \node (Y) [right of=SMT,align=center]  {};
  \node (Rad) [right of=Y,align=center] {$\mathcal{I}_\pm$\emph{radiative}\\\emph{phase space}};
  \draw[->] (QG) to node {$\hbar\rightarrow 0$} (LG);
  \draw[->] (LG) to node {$r\rightarrow\infty$} (Rad);
  \draw[->, dashed] (QG) to node [swap] {$r\rightarrow\infty$} (SMT);
  \draw[->] (Rad) to node [swap,align=center] {\emph{asymptotic}\\\emph{quantisation}} (SMT);
\end{tikzpicture}

\caption{\emph{The case for a quasi-local quantization of the gravitational field (QLQG).} The results of this research demonstrate that this diagram may not commute. }\label{FigMotvtn}  
\end{figure}

\section{Quasi-local graviton and the Barbero--Immirzi parameter}\label{sec2}
\subsection{Null initial data in terms of $SU(1,1)$ variables}
\paragraph{Basic remarks} In the following, we consider gravitational null initial data on a partial Cauchy surface, which is null (light-like). The resulting phase space describes the two radiative modes---non-linear gravitons \cite{AshtekarNullInfinity}---crossing the null surface and additional edge modes \cite{Balachandran:1994up,PhysRevD.51.632,Geiller:2017whh,Wieland:2017cmf,Wieland:2017zkf,Donnelly:2015hxa,Donnelly:2016auv,DonnellyGiddings2016,Gomes:2016mwl,Geiller:2017xad,Speranza:2017gxd,Takayanagi:2019tvn} at the co-dimension two boundaries of $\mathcal{N}$. If we exclude caustics, which are events on $\mathcal{N}$ where light rays would exit, the null surface $\mathcal{N}$ is turned naturally into a fibre bundle $(\mathcal{N},\mathcal{C},\mathrm{pr})$ consisting of the three-manifold $\mathcal{N}$ itself, a two-dimensional base manifold $\mathcal{C}$ and a projection $\mathrm{pr}:\mathcal{N}\rightarrow\mathcal{C}$. For any point $x$ on the base manifold, i.e.\ $x\in\mathcal{C}$, its pre-image $\mathrm{pr}^{-1}(x)$ defines a null generator, i.e.\ a light ray tangential to $\mathcal{N}$. Conversely, if $\mathrm{pr}_\ast:T\mathcal{N}\rightarrow T\mathcal{C}$ is the push-forward, any vector field $\ell^a\in T\mathcal{N}$ is null (light-like), if and only if $\mathrm{pr}_\ast \ell^a=0$. In the following, we simplify the problem further. We restrict ourselves to only such null surfaces that have the topology of a cylinder with two disconnected spacelike boundaries, $\mathcal{N}=[0,2\pi]\times{S}^2$, $\mathcal{C}_+=\{2\pi\}\times{S}^2$ and $\mathcal{C}_-=\{0\}\times[S^2]^{-1}$. That the time coordinate at the upper and lower boundary assumes the particular values $0$ and $2\pi$ was done to match common definitions of two-dimensional CFTs (conformal field theories). More about that in \hyperref[sec3]{Section 3}.
\smallskip 

Excluding caustics seems drastic. We will see below how this restriction takes on a new meaning at the quantum level. In our model, the base manifold is tessellated into $N$ faces. Each of these faces represents a fundamental atom of geometry carrying quantum numbers for the cross-sectional area, expansion and shear of the null generators of $\mathcal{N}$. In loop quantum gravity, we have a very similar non-perturbative quantisation of isolated horizons, see \cite{BarberoG.:2012ae,Ashtekar:2000eq}. 
 The exclusion of caustics is analogous to the preservation of particle number of the model. This observation suggests caustics can only arise in an improved model in which particle number is no longer conserved. This clearly suggests an avenue for future reseatch. 
In fact, it would very useful to consider a second quantisation of the present first-quantised model. In such a second-quantized approach, null generators can be excited out of a pre-geometric vacuum, i.e.\ a state representing no geometry at all. This viewpoint resonates with group field theory \cite{oriti,Freidel:2005qe,Oriti:2013aqa,Gielen:2024sxs} and loop quantum gravity \cite{Ashtekar:2021kfp,status,Rovelli:2014ssa,rovelli,alexreview,Engle2023,thiemann,ashtekar,Thiemann2023,Asante2023}, in which a generic solution to Einstein's equations may only arise from a highly excited state that is a coherent superposition of many atoms of space \cite{Oriti:2013jga}. We will comment more on this connection in the discussion below.

\paragraph{Dyadic co-basis on $\mathcal{N}$} Moving forward, we need to introduce some basics elements of the differential geometry of the null boundary. Taking the pull-back of the four-dimensional metric tensor to the null boundary, we obtain the signature $(0$$+$$+)$ three-metric
\begin{equation}
q_{ab}=\varphi^\ast_{\mathcal{N}}g_{ab}.\label{q-def}
\end{equation}
This metric can be diagonalised in terms of a one-form $m_a\in\Omega^1(\mathcal{N}:\C)$ and its complex conjugate $\bar{m}_a$ such that
\begin{equation}
q_{ab}=2 m_{(a}\bar{m}_{b)}.\label{q-dyad}
\end{equation}
 For any null vector $\ell^a\in T\mathcal{N}$, i.e.\ $\mathrm{pr}_\ast \ell^a=0$, we must then also have $m_a\ell^a=0$. Given $q_{ab}$, the one-form $m_a$ is unique up to $U(1)$ frame rotations sending $m_{a}$ into $\E^{\I\lambda}m_a$ with $\lambda$ being real. As explained in detail below, see also \cite{Wieland:2020gno,Wieland:2021vef}, these transformations give rise to a local gauge symmetry. The phase space for the $\gamma$-Palatini (Holst) action turns the corresponding charge, which would vanish for metric gravity, into the oriented area\footnote{Depending on the orientation, the $U(1)$ charge density can be positive or negative. To distinguish this observable from the geometrical area, i.e.\ $\mathring{\mathrm{Ar}}[\mathcal{S}]=\int_\mathcal{S} \di\sigma^1\di\sigma^2\sqrt{g(\partial_1,\partial_1)g(\partial_2,\partial_2)-g(\partial_1,\partial_2)^2}$, which is always greater or equal to zero, the $U(1)$ charge is sometimes referred to as the \emph{area-flux}, see \cite{Wieland:2017cmf,FernandoBarbero:2009ai}.} of the corners $\mathcal{C}_\pm$.  These charges are the integral of the boundary intrinsic two-form $\varepsilon\in\Omega^2(\mathcal{N}:\R)$ along $\mathcal{C}_\pm$, i.e.\ $\mathrm{Ar}[\mathcal{C}_\pm]=\int_{\mathcal{C}_\pm}\varepsilon$. The two-form $\varepsilon$ is simply the wedge product of the co-vectors $m_a$ and $\bar{m}_a$, i.e.\
\begin{equation}
\varepsilon_{ab}=-2\I m_{[a}\bar{m}_{b]}.
\end{equation}
In the following, we keep the abstract null bundle $(\mathcal{N},\mathcal{C},\mathrm{pr})$ fixed. In this way, the triple $(\mathcal{N},\mathcal{C},\mathrm{pr})$ becomes a universal reference structure shared between different spacetime geometries. This viewpoint is useful, because it puts a sort-of scaffolding around the otherwise structureless three-manifold $\mathcal{N}$. The triple $(\mathcal{N},\mathcal{C},\mathrm{pr})$ provides a natural \emph{causal order} \cite{Surya:2019ndm}, which is inherited from the physical geometry in the bulk. Fixing this structure is less restrictive than working on a fixed metric background. It is useful for us, because it simplifies the parametrization of physical fields. To understand how this can work in practice, consider the one-forms $(m_a,\bar{m}_a)$. To parametrise all possible one-forms $m_a$ compatible with $(\mathcal{N},\mathcal{C},\mathrm{pr})$, we equip the base manifold with an auxiliary (round) metric ${}^oq_{ab}$. With respect to this metric, we introduce standard spherical coordinates $(\cos\theta,\E^{\I\phi})$, or, which is the same thing, consider the stereographic complex coordinate
\begin{equation}
\zeta=\frac{1}{\tan{\frac{\theta}{2}}}\E^{\I\phi}.\label{zeta-def}
\end{equation}
This coordinate naturally extends through $\zeta':=\zeta\circ\mathrm{pr}$ into a stereographic coordinate along $\mathcal{N}$. From now on, we do not distinguish the two, i.e.\ $\zeta'\equiv\zeta$. Given these coordinates, we introduce the following reference co-dyad on $\mathcal{N}$, namely
\begin{equation}
{}^{o}m=\sqrt{2}\frac{\di\zeta}{1+|{\zeta}|^2},\qquad{}^{o}\bar{m}=\sqrt{2}\frac{\di\bar{\zeta}}{1+|{\zeta}|^2}.
\end{equation}
The corresponding area element is
\begin{equation}
d^2v_o=-\I {}^om\wedge{}^o\bar{m}=\sin\vartheta\,\di\vartheta\wedge\di\varphi.\label{d2vo-def}
\end{equation}
For the physical co-dyad $(m_a,\bar{m}_a)$, we can then use the following matrix parametrisation
\begin{equation}
\begin{pmatrix}
m_a\\\bar{m}_a
\end{pmatrix}=\Omega\,
\begin{pmatrix}
\ou{S}{0}{0}&\ou{S}{0}{1}\\
\ou{S}{1}{0}&\ou{S}{1}{1}
\end{pmatrix}
\begin{pmatrix}
{}^{o}m_a\\{}^{o}\bar{m}_a
\end{pmatrix}\equiv\Omega\,S\cdot\begin{pmatrix}
{}^{o}m_a\\{}^{o}\bar{m}_a
\end{pmatrix}.
\end{equation}
In here, $\Omega:\mathcal{N}\rightarrow\R_>$ is the conformal factor and the matrix $S$ has determinant one. The condition that the one-form $\bar{m}_a$ is the complex conjugate of $m_a$, implies that the two-by-two matrix $S$ takes, in fact, values in $SU(1,1)$. The intuitive meaning of this parametrisation is clear. The conformal factor determines the overall scale, the $SU(1,1)$ group element determines the local shape modes. The idea of promoting such shape modes into quantum observables clearly resonates with \emph{shape dynamics}, which is an approach to quantum gravity based on similar conformal methods applied to initial data on spacelike  rather than null hypersurfaces \cite{shapebook,Gomes:2010fh}.

\paragraph{Shear and expansion} Since the metric $q_{ab}$ is degenerate, there is no unique metric compatible derivative available (on any two-dimensional cut, there is). On the other hand, the Lie derivative $\mathcal{L}_\xi$ for $\xi^a\in \mathcal{N}$ is always available. In the following, the Lie derivative along the null generators $\ell^a\in TN:\mathrm{pr}_\ast\ell^a=0$ plays an important role. In defines a natural notion of time evolution along the null generators. Consider thus the Lie derivative of the dyadic basis. We obtain
\begin{equation}
\mathcal{L}_\ell m_a=\frac{1}{2}\vartheta_{(\ell)}m_a+\I\varphi_{(\ell)}m_a+\sigma_{(\ell)}\bar{m}_a.\label{Lie-m}
\end{equation}
In here, $\vartheta_{(\ell)}$ is the expansion, $\varphi_{(\ell)}$ is the time component of a $U(1)$ connection and $\sigma_{(\ell)}$ is the shear. The expansion is the time derivative of the conformal factor
\begin{equation}
\vartheta_{(\ell)}=\Omega^{-1}\mathcal{L}_\ell\Omega.
\end{equation}
On the other hand, $\varphi_{(\ell)}$ and $\sigma_{(\ell)}$ are componets of the $\mathfrak{su}(1,1)$-valued Maurer--Cartan form $\di SS^{-1}$ on $\mathcal{N}$. In fact,
\begin{equation}
\mu_{(\ell)}:=(\mathcal{L}_\ell S)S^{-1}=\varphi_{(\ell)}J+\sigma_{(\ell)}\bar{X}+\bar{\sigma}_{(\ell)}X.
\end{equation}
In here, we introduced the following standard basis of the Lie algebra,
\begin{equation}
J=\begin{pmatrix}\I &0\\0&-\I\end{pmatrix},\qquad
X=\begin{pmatrix}0 &1\\0&0\end{pmatrix},\qquad
\bar{X}=\begin{pmatrix}0 &0\\1&0\end{pmatrix}\label{su1-1-basis}
\end{equation}
These matrices together satisfy the standard $SU(1,1)$ Pauli relations
\begin{equation}
\left.
\begin{split}
J^2&=-\bbvar{1},\\
X^2&=\bar{X}^2=0,
\end{split}\qquad
\begin{split}
J X &=-XJ=-\I X,\\
X\bar{X}&=\tfrac{1}{2}(\bbvar{1}+\I J),
\end{split}\qquad
\begin{split}
J \bar{X} &=-\bar{X}J=\I \bar{X},\\
\bar{X}X&=\tfrac{1}{2}(\bbvar{1}-\I J),
\end{split}\quad
\right\}
\end{equation}
where $\bbvar{1}=\left(\begin{smallmatrix}1&0\\0&1\end{smallmatrix}\right)$ is the identity matrix. We then also have the commutators $[J,X]=-2\I X$, $[J,\bar{X}]=+2\I X$, $[X,\bar{X}]=\I J$ and trace relations $\mathrm{Tr}(J^2)=-2$, $\mathrm{Tr}(X\bar{X})=1$.

\subsection{Non-affinity, Raychaudhuri equation and choice of clock}
\paragraph{Affinity of null generators}Each of the null generators of $\mathcal{N}$ is a geodesic with respect to the four-dimensional metric $g_{ab}$ in the bulk. If $\ell^a\in T\mathcal{N}$ is a null vector, i.e.\ $\mathrm{pr}_\ast \ell^a=0$, we then have
 \begin{equation}
 \ell^bD_b\ell^a=\kappa_{(\ell)}\ell^a,\label{lDl-kappa-def}
 \end{equation}
where $D_a$ is the pull-back of the Levi-Civita covariant derivative from the bulk to the boundary.\footnote{Since the metric $q_{ab}$ is degenerate, there is no natural metric compatible and torsion-less covariant derivative on $\mathcal{N}$. The two natural derivatives available on $\mathcal{N}$ are the Lie derivative $\mathcal{L}$, which does not require a metric, and the pull-back $D_a$ of the covariant derivative from the bulk. The $D_a$-derivative does not act on the tensor bundle $T\mathcal{N}$. Instead it acts on elements of the four-dimensional tensor bundle $T\mathcal{M}$ restricted to $\mathcal{N}$. It is the tensorial and null-surface analogue of the selfdual Ashtekar connection \cite{ashtekar}, which is defined on a spacelike hypersurface and depends on both its extrinsic and intrinsic geometry. In here, the situation is similar, $D_a$ depends  on $q_{ab}$ but also the embedding of $\mathcal{N}$ into spacetime $(\mathcal{M},g_{ab})$.} A dilation sends $\ell^a$ into $\ell'^a=\E^{-\lambda} \ell^a$ for $\lambda:\mathcal{N}\rightarrow\R$. Under such a map, $\kappa_{(\ell)}$ transforms as
\begin{equation}
\kappa_{(\ell')}=\E^{-\lambda}(\kappa_{(\ell)}-\mathcal{L}_\ell\lambda)\quad\text{for}\quad \ell'^a=\E^{-\lambda} \ell^a,\label{kappa-dilat}
\end{equation}
where $\mathcal{L}_\ell\lambda=\ell^aD_a\lambda$.
\paragraph{Raychaudhuri equation and choice of clock} The only constraint that the vacuum Einstein equations impose on the variables $(S,\Omega,\kappa_{(\ell)})$ is the Raychaudhuri equation
\begin{equation}
\mathcal{L}_\ell\vartheta_{(\ell)}= \ell^aD_a\vartheta_{(\ell)}=\kappa_{(\ell)}\vartheta_{(\ell)}-\frac{1}{2}\vartheta_{(\ell)}^2-2\sigma_{(\ell)}\bar{\sigma}_{(\ell)}.\label{Raychudhri-eq1}
\end{equation}
To continue, we choose a (future pointing) null vector field $\mathring{\ell}^a$ such that
\begin{equation}
{\kappa}_{\smallcirc}:=\kappa_{(\mathring{\ell})}=-\frac{1}{2}\vartheta_{(\mathring{\ell})}.\label{kappa-def}
\end{equation}
This condition can always be satisfied using a dilation of the null generators, see \eref{kappa-dilat}. To simplify our notation moving forward, we write
\begin{equation}
(\varphi_{\smallcirc},{\vartheta}_{\smallcirc},{\sigma}_{\smallcirc}):=(\varphi_{(\mathring{\ell})},\vartheta_{(\mathring{\ell})},\sigma_{(\mathring{\ell})}).\label{circ-spincoeff}
\end{equation}

Next, we integrate this vector field obtaining a clock variable $\mathcal{U}:\mathcal{N}\rightarrow\R$, $\mathring{\ell}^a\partial_a\mathcal{U}=1$. In this way, we can complete the two-dimensional $\zeta$-coordinates introduced above, see \eref{zeta-def}, into a three-dimensional coordinate system $(\mathcal{U},\zeta,\bar{\zeta})$ on $\mathcal{N}$. In particular, the vector field
\begin{equation}
\mathring{\ell}^a=\partial^a_{\mathcal{U}}\label{l-U-def}
\end{equation}
is null. Notice that the condition \eref{kappa-def} does not fix the clock variable uniquely. We remove the residual freedom by imposing the boundary conditions
\begin{equation}
\mathcal{U}\big|_{\mathcal{C}_-}=0,\qquad \mathcal{U}\big|_{\mathcal{C}_+}=2\pi.\label{U-bndrycond}
\end{equation}
A time function $\mathcal{U}:\mathcal{N}\rightarrow\R$ that satisfies (\ref{kappa-def}, \ref{U-bndrycond}) defines what we call a \emph{teleological clock} on $\mathcal{N}$. With respect to this clock, the Raychaudhuri equation \eref{Raychudhri-eq1} becomes a second-order differential equation for the conformal factor
\begin{equation}
\frac{\di^2}{\di\mathcal{U}^2}\Omega^2=-2{\sigma}_\smallcirc{\bar{\sigma}_\smallcirc}\Omega^2.\label{Raychudhri-eq2}
\end{equation}
Notice that for vanishing shear, the solution simplifies. If $\sigma_\smallcirc=0$, the conformal factor becomes a linear function of $\mathcal{U}$, i.e.\
\begin{equation}
\sigma_{\smallcirc}=0\Leftrightarrow \Omega^2(\mathcal{U},\zeta,\bar{\zeta})=\frac{1}{2\pi}\left(\Omega^2_+(\zeta,\bar{\zeta})\mathcal{U}+\Omega^2_-(\zeta,\bar{\zeta})(2\pi-\mathcal{U})\right),
\end{equation}
where $\Omega^2_\pm$ are the initial conditions for the conformal factor at the upper (lower) cut.
\subsection{Oscillator representation of the kinematical phase space}
\paragraph{Kinematical phase space} The starting point for our construction  is the pre-symplectic potential $\Theta_{\mathcal{N}}$ for the $\gamma$-Holst--Palatini action on the light-like boundary $\mathcal{N}$. We have included a concise derivation  in the \hyperref[appdx]{Appendix} to this paper, for a more comprehensive analysis see \cite{Wieland:2021vef,Wieland:2017cmf}.\smallskip

Equation \eref{kappa-def} implies that the teleological clock $\mathcal{U}:\mathcal{N}\rightarrow [0,2\pi]$ is a functional on phase space, hence $\bbvar{d}\mathcal{U}\neq 0$, where $\bbvar{d}$ is the exterior derivative on field space. To parametrize the time dependence of the various modes on phase space, it is practical to  consider an auxiliary unphysical clock, which is fixed once and for all, i.e.\ $\bbvar{d}u=0$. The $u$-coordinate defines a good clock, if it grows monotonically relative to $\mathcal{U}$, which is the same as to say\footnote{In the following,  $\dot{O}\equiv\partial_uO$ for all functionals $O$ on phase space.}
\begin{equation}
\partial_u\mathcal{U}\equiv \dot{\mathcal{U}}>0.
\end{equation}
This suggests to introduce the parametrization
\begin{equation}
 \dot{\mathcal{U}}=\E^\chi.\label{chi-def}
\end{equation}
In our analysis $\chi$ (chronoton) is one of our dynamical fields on phase space. It is a sort of lapse function that determines the deviation between the auxiliary $u$-coordiante from the teleological clock $\mathcal{U}$. If $\chi=0$, the two clocks coincide.\smallskip

In terms of the fundamental boundary fields,  the pre-symplectic potential for the $\gamma$-Holst action is
\begin{align}\nonumber
\Theta_{\mathcal{N}}&=-\frac{1}{8\pi G}\int_{\mathcal{C}_+} d^2v_o\,\Omega^2\bbvar{d}\lambda+\frac{1}{8\pi G}\int_{\mathcal{N}}d^3v_o\,\frac{\di}{\di u}\left[\Omega^2\right]\bbvar{d}\chi+\\
&\quad+\frac{1}{8\pi\gamma G}\int_{\mathcal{N}}d^3v_o\,\Omega^2\bbvar{d}\varphi+\int_{\mathcal{N}}d^3v_o\,\mathrm{Tr}\left(\Pi(\bbvar{d}S)S^{-1}\right).\label{theta-N1}
\end{align}
For a derivation of this expression, see \eref{app:theta-N4} below and reference \cite{Wieland:2021vef,Wieland:2017cmf}. In here, $\lambda$ is a gauge parameter responsible for the dilatations of the null normal, see \eref{kappa-dilat} and \eref{app:null-dilat} and $d^3v_o$ is the auxiliary volume element,
\begin{equation}
d^3v_o=\di u\wedge d^2v_o=-\I\,\di u\wedge{}^om\wedge{}^o\bar{m}.\label{d3vo-def}
\end{equation}
The fundamental variables that appear in \eref{theta-N1} are not independent from each other. There are  constraints among them. They can be inferred by first computing the $\mathfrak{su}(1,1)$ Lie algebra element
\begin{equation}
\dot{S}S^{-1}=:\varphi J+\sigma\bar{X}+\bar{\sigma}X=:\mu.\label{LieS}
\end{equation}
Our earlier definitions \eref{Lie-m} and \eref{circ-spincoeff} imply
\begin{equation}
(\varphi_\smallcirc,\vartheta_\smallcirc,\sigma_\smallcirc)=\E^{-\chi}(\varphi,\vartheta,\sigma).
\end{equation}
In addition, the $\mathfrak{su}(1,1)$-valued momentum satisfies the decomposition
\begin{equation}
\Pi = LJ+c\bar{X}+\bar{c}X. \label{Pi-def}
\end{equation}
In here, the individual components are constrained to satisfy
\begin{align}
L&=-\frac{1}{16\pi\gamma G}\frac{\di}{\di u}\Omega^2,\label{L-def}\\
c&=-\frac{1}{8\pi\gamma G}(\gamma+\I)\Omega^2\sigma,\label{c-def}
\end{align}
see also \eref{app:Lcons} and \eref{app:ccons} below, in which all quantities are expressed with respect to the physical $\mathcal{U}$ clock rather than the auxiliary variable $u$, c.f.\ \eref{app:spincoeff}. In short, equation \eref{LieS} and \eref{Pi-def} imply a constraint between the $SU(1,1)$-valued configuration variable $S$ and the $\mathfrak{su}(1,1)$-valued momentum $\Pi$. In addition, there is a constraint between the momentum conjugate to $\chi$ and the diagonal component of $\Pi$. Finally, we have to impose the Raychaudhuri equation \eref{Raychudhri-eq2}, which turns into a constraint on phase space. Before going into the discussion of the constraint algebra and the choice of canonical variables, let us add a few remarks about an implicit $U(1)$ symmetry in the model.

\paragraph{$U(1)$ holonomy} The co-frame $m_a$ is charged under a $U(1)$ gauge symmetry, sending $m_a$ into $\E^{\I\lambda}m_a$, with $\lambda$ being a \emph{local} $\R$-valued function on $\mathcal{N}$. Under such a map, $S\in SU(1,1)$ transforms as $S\rightarrow \E^{J\lambda}S$. In addition, $\varphi$ transforms as the time component of a $U(1)$ connection and $\sigma$ is sent into $\sigma\E^{-2\I\lambda}$, reflecting the spin-2 nature of the radiative modes of the gravitational field. Since $\varphi$ is a connection component, it seems appropriate to consider the corresponding $U(1)$ holonomy $\E^{\I\Delta}$, where
\begin{equation}
\Delta(u,\zeta,\bar\zeta) = \int_0^u\di u'\,\varphi(u',\zeta,\bar{\zeta})\,\mathrm{mod} 2\pi.\label{Delt-def}
\end{equation}
This function will play an important role in the definition of the canonical variables.

\paragraph{Oscillator representation of the area modes} The $U(1)$ angle \eref{Delt-def} takes values in a compact interval. At the quantum level, this means that the conjugate variables, which plays the role of the area has a discrete spectrum. This becomes particularly simple to understand in terms of certain oscillator type variables. Consider, for this purpose, the $U(1)$ part of the symplectic potential, which contains the functional differential $\bbvar{d}\varphi$ of $\varphi$. Going back to \eref{theta-N1} and taking into account $\dot\Delta=\varphi$, $\bbvar{d}(\dot\Delta)=\tfrac{\di}{\di u}\bbvar{d}\varphi$, which is a consequence of $\bbvar{d}u=0$, we obtain
\begin{align}\nonumber
-\int_{\mathcal{C}_+} d^2v_o\,\Omega^2\bbvar{d}\lambda+&\gamma^{-1}\int_{\mathcal{N}}d^3v_o\,\Omega^2\bbvar{d}\varphi=\\
&=\gamma^{-1}\int_{\mathcal{C}_+} d^2v_o\,\Omega^2\bbvar{d}(\Delta-\gamma\lambda)-\gamma^{-1}\int_{\mathcal{N}}d^3v_o\,\frac{\di}{\di u}\left[\Omega^2\right]\bbvar{d}\Delta.
\end{align}
This equation suggests to introduce the following oscillator variables.\smallskip

\noindent At the upper boundary, we introduce the edge mode oscillators
\begin{equation}
a(\zeta,\bar{\zeta})=\frac{1}{\sqrt{8\pi\gamma G}}\left|\Omega_+\right|\E^{-\I(\Delta_+-\gamma\lambda)},\label{a-def}
\end{equation}
where $\Omega_+=\Omega\big|_{\mathcal{C}_+}$ and $\Delta_+=\Delta\big|_{\mathcal{C}_+}$ are the restriction of the conformal factor $\Omega$ and the $U(1)$ angle to the upper cut $\mathcal{C}_+\subset\mathcal{N}$.\smallskip

\noindent  Along $\mathcal{N}$, we introduce the oscillators
\begin{equation}
b = \frac{1}{\sqrt{8\pi\gamma G}}\sqrt{\left|\frac{\di}{\di u}\Omega^2\right|}\E^{\I r\Delta},\label{b-def}
\end{equation}
where
\begin{equation}
r=\operatorname{sgn}\left(\frac{\di}{\di u}\Omega^2\right)\label{r-def}.
\end{equation}
The sign of $\Omega^{-1}\dot{\Omega}$ is a discrete observable that tells us whether the null surface is expanding or contracting. At the classical level, $r$ splits the phase space into two regions separated by the $r=0$ surface that characterizes isolated horizons \cite{Ashtekar:2000eq,isohorizon,Ashtekar:1999wa,Ashtekar:aa,BarberoG.:2012ae,DiazPoloPranzetti,Ashtekar:2001is,Engle:2010kt}. At the quantum level, $r$ labels different components of the wavefunction---think of a particle on the real line. We can split the classical phase space $\R^2\ni (p,q)$ into two half spaces corresponding to $r=\operatorname{sgn}(p)$. This amounts to treating  left-moving and right-moving configurations separately.  Quantum mechanically, this is realized by splitting the Hilbert space $L^2(\R,\di p)$ into a direct sum of wavefunctions that are confined to the upper and lower half-planes, i.e.\ $L^2(\R,\di p)=L^2(\R_>,\di p)\otimes L^2(\R_<,\di p)$. In our case, special junction conditions need to be imposed for the $r=0$ configurations. These will be studied elsewhere.  In the following, we restrict ourselves to the $r\leq0$ case, thus considering an infalling null surface---e.g.\ a portion of a large null cone sufficiently close to future null infinity such that no caustics can form (a version of \emph{finite infinity}, see \cite{Ellis1984}).\smallskip

In either case, the kinematical Poisson brackets (i.e.\ prior to imposing the constraints) among these oscillators are then given by\footnote{In what follows, vanishing Poisson brackets are not stated explicitly.}
\begin{align}
\{a(\zeta,\bar{\zeta}),\bar{a}(\zeta',\bar{\zeta}')\}&=\I\, \delta_{\mathcal{C}_+}(\zeta-\zeta',\bar{\zeta}-\bar{\zeta}'),\label{a-oprts}\\
\{b(u,\zeta,\bar{\zeta}),\bar{b}(u',\zeta',\bar{\zeta}')\}&=\I\, \delta_{\mathcal{N}}(u-u',\zeta-\zeta',\bar{\zeta}-\bar{\zeta}'),\label{b-oprts}
\end{align}
where the two Dirac delta distributions are normalized with respect to the measures $d^2v_o$ and $d^3v_o$, see \eref{d2vo-def} and \eref{d3vo-def}.

\paragraph{Heisenberg representation of the $SU(1,1)$ sector of phase space} In the same way as we introduced the $a$, $\bar{a}$, $b$, $\bar{b}$ oscillators, we can build a Heisenberg-type of representation of the $SU(1,1)$ sector of phase space. This was introduced first for impulsive null initial data in \cite{Wieland:2024dop}. Here, we generalize the construction to the continuum.\smallskip

To start out, we introduce a few basic elements about $SU(1,1)$. In what follows, we utilize the standard spinor (twistor) notation from  \cite{penroserindler}. Abstract tensor $A,B,C\dots$ denote the sections of a two-dimensional vector bundle on $\mathcal{N}$ equipped with two $SU(1,1)$-invariant bilinear forms $\epsilon_{AB}$ and $\eta_{A\bar{A}}$, where $\epsilon_{AB}=-\epsilon_{BA}$ is the skew-symmetric Levi-Civita spinor and $\eta_{A\bar{A}}$ is the signature $(-,+)$ hermitian inner product.. Indices $\bar{A},\bar{B},\bar{C},\dots$ belong to the complex conjugate representation. We can then realize $SU(1,1)$ as the group of complex $2\times 2$ matrices $\ou{S}{A}{B}$,  that preserve the two bilinears,
\begin{align}
\epsilon_{CD}\ou{S}{C}{A}\ou{S}{D}{B}=\epsilon_{AB},\qquad \eta_{B\bar{B}}\ou{S}{B}{A}\ou{\bar{S}}{\bar{B}}{\bar{A}}=\eta_{A\bar{A}}.\label{SU1-1-invrncs}
\end{align}
 We then also have the inverse Levi-Civita tensor $\epsilon^{AB}$ that satisfies $\epsilon^{AC}\epsilon_{BC}=\delta^A_B$. The $\C^2$-indices $A,B,C,\dots$ are raised and lowered by the Levi-Civita tensors.\footnote{Conventions are $\xi^A=\epsilon^{AB}\xi_B$, $\xi_A=\xi^{B}\epsilon_{BA}$. Notice $\xi^A\eta_A=-\xi_A\eta^A$.} 
In this notation, we obtain e.g.\ $\overline{\eta_{A\bar{A}}}=\eta_{A\bar{A}}$, $\eta^{A\bar A}=\epsilon^{AB}\bar{\epsilon}^{\bar{A}\bar{B}}\eta_{B\bar{B}}$, $\eta^{A\bar{C}}\eta_{B\bar{C}}=-\delta^A_B$.\smallskip

The $\mathfrak{su}(1,1)$ momentum $\ou{\Pi}{A}{B}$, see \eref{Pi-def}, is traceless. Any traceless $2\times 2$ complex matrix can be diagonalised using its two eigenspinors. Call them $\pi_A$ and $\omega_A$ with eigenvalues $\pm\pi_A\omega^A$. We obtain
\begin{equation}
\Pi_{AB}=\pi_{(A}\omega_{B)}\equiv\frac{1}{2}\left(\pi_{A}\omega_{B}+\pi_{B}\omega_{A}\right).\label{eigen-spinrs}
\end{equation}
The components of $\ou{\Pi}{A}{B}$, which are related to shear and expansion, see \eref{L-def} and \eref{c-def}, can be thus parametrized as
\begin{align}
L&=-\frac{1}{2}\mathrm{Tr}(\Pi J)\equiv-\frac{1}{2}\ou{J}{A}{B}\ou{\Pi}{B}{A}=\frac{1}{2}J^{AB}\pi_A\omega_B,\label{L-piom}\\
c&=\mathrm{Tr}(X\Pi)\equiv\ou{X}{A}{B}\ou{\Pi}{B}{A}=-X^{AB}\pi_A\omega_B.\label{c-piom}
\end{align}
For $\ou{\Pi}{A}{B}$ to lie in $\mathfrak{su}(1,1)$ rather than $\mathfrak{sl}(2,\C)$, we need to satisfy an additional condition, namely the \emph{reality conditions}
\begin{equation}
\ou{\Pi}{A}{B}=\eta^{A\bar{A}}\eta_{B\bar{B}}\ou{\bar{\Pi}}{\bar{B}}{\bar{A}}.\label{realcond}
\end{equation}
This condition translates into a condition for the eigenspinors $(\pi_A,\omega^A)$ themselves. There are only two cases, which can be distinguished by the sign of the $SU(1,1)$-invariant $\Tr(\Pi^2)=-\Pi_{AB}\Pi^{AB}=\frac{1}{2}(\pi_A\omega^A)^2$. For reasons that should become clear below, we shall say that these two cases represent \emph{infra-Planckian} and \emph{ultra-Planckian} modes, to be distinguished as:\footnote{It would be confusing to call these modes \emph{trans-Planckian}, because this term has already a very specific meaning in black hole physics and cosmology. Therefore, we use here a slightly different terminology.} 
\begin{itemize}
\item \emph{Infra-Planckian modes, $L^2\geq c\bar{c}$.} In this case, $\pi_A\omega^A$ is imaginary. The reality conditions \eref{realcond} for the $\mathfrak{su}(1,1)$-valued momentum variable $\Pi$ translate into
\begin{equation}
\pi_A=\pm\I\eta_{A\bar A}\bar{\omega}^{\bar{A}},\qquad\omega_A=\pm\I\eta_{A\bar{A}}\bar{\pi}^{\bar{A}}.\label{realcond1}
\end{equation}
The relative sign matches the discrete $r$-observable introduced in \eref{r-def}, such that $\pi_A=\I\,r\,\eta_{A\bar A}\bar{\omega}^{\bar{A}}$.
\item \emph{Ultra-Planckian modes, $L^2\leq c\bar{c}$.} In this case, $\pi_A\omega^A$ is real.  The reality conditions \eref{realcond} for the $\mathfrak{su}(1,1)$-valued momentum variable $\Pi$  translate into
\begin{equation}
{\pi}_{A}=\eta_{A\bar A}\bar{\pi}^{\bar{A}},\qquad{\omega}_{A}=\eta_{A\bar{A}}\bar{\omega}^{\bar{A}}.\label{realcond2}
\end{equation}
\end{itemize}
The same conditions also apply to $\utilde{\pi}_A$ and $\utilde{\omega}^A$. These \emph{reality conditions} constrain the inner product of the quantum theory. The physical significance of this observation will be discussed below.\smallskip

The eigenspinors $(\pi_A,\omega_A)$, can be now used as new canonical variables for the $SU(1,1)$ sector themselves. Taking into account that for any $\ou{S}{A}{B}\in SU(1,1)$, we can write the matrix elements of $S^{-1}$ as $\ou{[S^{-1}]}{A}{B}=-\uo{S}{B}{A}$, we obtain
\begin{align}\nonumber
\int_{\mathcal{N}}d^3v_o\Tr\left(\Pi(\bbvar{d}S)S^{-1}\right)&\equiv\int_{\mathcal{N}}d^3v_o\,\ou{\Pi}{A}{B}\bbvar{d}(\ou{S}{B}{C})\ou{[S^{-1}]}{C}{A}=\int_{\mathcal{N}}d^3v_o\,\Pi_{AB}\bbvar{d}\ou{S}{B}{C}S^{AC}=\\\nonumber
&=\int_{\mathcal{N}}d^3v_o\left(\pi_A\bbvar{d}(\omega_B\ou{S}{B}{C})S^{AC}-\pi_A\ou{S}{B}{C}S^{AC}\bbvar{d}\omega_B\right)=\\
&=\int_{\mathcal{N}}d^3v_o\left(\pi_A\bbvar{d}\omega^A-\utilde{\pi}_A\bbvar{d}\utilde{\omega}^A\right).
\end{align}
where we used the fact that $S^{-1}S$ is traceless to drop the symmetrization over the indices $A$ and $B$. In here, we introduced
\begin{align}
\utilde{\omega}^A=\ou{[S^{-1}]}{A}{B}\omega^B,\qquad\utilde{\pi}^A=\ou{[S^{-1}]}{A}{B}\pi^B.\label{t-piom-def}
\end{align}
The fundamental kinematical Poisson brackets for these variables are
\begin{align}
\left\{\pi_A(u,\zeta,\bar{\zeta}),\omega_B(u',\zeta,\bar{\zeta})\right\}&=+\epsilon_{AB}\delta_{\mathcal{N}}(u-u',\zeta-\zeta',\bar{\zeta}-\bar{\zeta}'),\label{piom1}\\
\left\{\utilde{\pi}_A(u,\zeta,\bar{\zeta}),\utilde{\omega}_B(u',\zeta,\bar{\zeta})\right\}&=-\epsilon_{AB}\delta_{\mathcal{N}}(u-u',\zeta-\zeta',\bar{\zeta}-\bar{\zeta}'),\label{piom2}
\end{align}

\paragraph{Kinematical symplectic potential for the oscillators} Before continuing, let us briefly summarize. Starting from the $\gamma$-Holst action, and evaluating the corresponding symplectic potential on a null  boundary, see \eref{theta-N1}, \eref{app:theta-N3} and references \cite{Wieland:2021vef,Wieland:2017cmf}, we introduced new canonical variables. As far as the kinematical phase space is concerned, instead of \eref{theta-N1}, we can then equally work with the symplectic potential
\begin{align}
\nonumber
\Theta_{\mathcal{N}}&=\frac{1}{2\I}\int_{\mathcal{C}_+}d^2v_o\left(a\bbvar{d}\bar a-\CC\right)+\\
&\nonumber\quad+\int_{\mathcal{N}}d^3v_o\,p_\chi\bbvar{d}\chi+\frac{1}{2\I}\int_{\mathcal{N}}d^3v_o\left(b\bbvar{d}\bar b-\CC\right)+\\
&\quad+\frac{1}{2}\int_{\mathcal{N}}d^3v_o\left(\pi_A\bbvar{d}\omega^A+\omega_A\bbvar{d}\pi^A-\utilde{\pi}_A\bbvar{d}\utilde{\omega}^A-\utilde{\omega}_A\bbvar{d}\utilde{\pi}^A\right).\label{theta-N2}
\end{align}
The resulting canonical Poisson brackets are given in \eref{a-oprts}, \eref{a-oprts}, and \eref{piom1}, \eref{piom2}. Besides the canonical Poisson brackets between the oscillators $a$, $\bar{a}$, $b$, $\bar{b}$ and the Heisenberg variables $(\pi_A,\omega^A;\utilde{\pi}_A,\utilde{\omega}^A)$, we also have the fundamental brackets between $\chi$ (chronoton) and its conjugate momentum on the kinematical phase space
\begin{equation}
\left\{p_\chi(u,\zeta,\bar{\zeta}),\chi(u',\zeta',\bar{\zeta}')\right\}=\delta_{\mathcal{N}}(u-u',\zeta-\zeta',\bar{\zeta}-\bar{\zeta}')\label{p-chi}
\end{equation}
Physical configurations define a symplectic submanifold characterized by a set of first-class and second-class constraints. The constraints consist of the Raychaudhuri constraint \eref{Raychudhri-eq2} and further constraints, to be inferred from \eref{L-def} and \eref{c-def}. The constraints and their algebra will be the subject of the rest of this section. In the next subsection, we discuss the first-class constraints and their algebra. \hyperref[sec2.5]{Section 2.5} is about the second-class constraints.
\subsection{First-class constraints and their algebra}
\paragraph{Hamilton constraint} Consider a generic null vector field on $\mathcal{N}$ and parameterize it in terms of the unphysical null time coordinate $u\in[0,\pi]$ and a lapse function $N(u,\zeta,\bar{\zeta}')$. We thus write
\begin{equation}
\xi^a_N=N(u,\zeta,\bar{\zeta}')\partial^a_u.
\end{equation}
It is now easy to check that the Lie derivative $\mathcal{L}_{\xi_N}$ acts on the fundamental kinematical variables as follows
\begin{align}
\mathcal{L}_{\xi_N}[b]&=N\dot{b}+\frac{1}{2}\dot{N}b,\label{Lxi-b}\\
\mathcal{L}_{\xi_N}[\bar{b}]&=N\dot{\bar{b}}+\frac{1}{2}\dot{N}\bar{b},\label{Lxi-bbar}\\
\mathcal{L}_{\xi_N}[\chi]&=N\dot{\chi}+\dot{N},\label{Lxi-chi}\\
\mathcal{L}_{\xi_N}[p_\chi]&=\frac{\di}{\di u}(Np_\chi),\label{Lxi-pchi}\\
\mathcal{L}_{\xi_N}[S]&=N\dot{S},\label{Lxi-S}\\
\mathcal{L}_{\xi_N}[\Pi]&=N\dot{\Pi}+\dot{N}\Pi.\label{Lxi-Pi}
\end{align}
If we replace $(\Pi,S)\in\mathfrak{su}(1,1)\times SU(1,1)$ by the Heisenberg variables $(\pi_A,\omega^A;\utilde{\pi}_A,\utilde{\omega}^A)$, we obtain
\begin{align}
\mathcal{L}_{\xi_N}\begin{pmatrix}\omega^A\\\pi_A\end{pmatrix}&=\begin{pmatrix}N\dot{\omega}^A+\tfrac{1}{2}\dot{N}\omega^A\\N\dot{\pi}_A+\tfrac{1}{2}\dot{N}\pi_A\end{pmatrix},\label{Lxi-piom}\\
\mathcal{L}_{\xi_N}\begin{pmatrix}\utilde{\omega}^A\\\utilde{\pi}_A\end{pmatrix}&=\begin{pmatrix}N\dot{\utilde{\omega}}^A+\tfrac{1}{2}\dot{N}\utilde{\omega}^A\\N\dot{\utilde{\pi}}_A+\tfrac{1}{2}\dot{N}\utilde{\pi}_A\end{pmatrix}.\label{Lxi-t-piom}
\end{align}
Consider now the corresponding generator on phase space for a smearing funtcion that vanishes at the upper and lower cuts, i.e.\
\begin{equation}
N\big|_{\mathcal{C}_\pm}=0.
\end{equation}
Going back to \eref{theta-N1} and \eref{LieS} and \eref{Pi-def}, \eref{L-def}, \eref{c-def}, we obtain the Hamilton constraint
\begin{align}\nonumber
H[N]:=\Theta_{\mathcal{N}}(\mathcal{L}_{\xi_N})&=\frac{1}{8\pi G}\int_{\mathcal{N}}d^3v_o\left[\frac{\di}{\di u}\Omega^2(N\dot{\chi}+\dot{N})\right.+\\\nonumber
&\qquad\left.-\frac{1}{\gamma}N\frac{\di}{\di u}\Omega^2\dot{\Delta}+8\pi GN\operatorname{Tr}\left(\Pi\dot{S}S^{-1}\right)\right]=\\\nonumber
&=\frac{1}{8\pi G}\int_{\mathcal{N}}d^3v_o\left[\frac{\di}{\di u}\Omega^2\E^{-\chi}\frac{\di}{\di u}\left(\E^\chi N\right)-2N\Omega^2\sigma\bar{\sigma}\right]=\\\nonumber
&=\frac{1}{8\pi G}\int_{\mathcal{N}}d^3v_o\,\E^{\chi}\left[\frac{\di}{\di \mathcal{U}}\Omega^2\E^{-\chi}\frac{\di}{\di \mathcal{U}}\left(\E^\chi N\right)-2N\E^{\chi}\Omega^2\sigma_\smallcirc\bar{\sigma}_\smallcirc\right]=\\
&=-\frac{1}{8\pi G}\int_{\mathcal{N}}d^3v_o\,N\E^{\chi}\left[\frac{\di^2}{\di \mathcal{U}^2}\Omega^2+2\Omega^2\sigma_\smallcirc\bar{\sigma}_\smallcirc\right]=0,
\end{align}
where the last step involved a partial integration. Upon imposing the Raychaudhuri equation \eref{Raychudhri-eq2}, the generator of reparametrisations of the clock variable, i.e.\ $H[N]:=\Theta_{\mathcal{N}}(\mathcal{L}_{\xi_N})$ vanishes as a constraint on phase space. Explicitly,
\begin{equation}
\forall N:\mathcal{N}\rightarrow\R:N\big|_{\mathcal{C}_\pm}=0:H[N]=0.\label{H-cons1}
\end{equation}
Let uns now express this constraint in terms of the $b$, $\bar{b}$ oscillators, the chronoton modes $p_\chi$, $\chi$  and the $SU(1,1)$ oscillators. A short calculation gives
\begin{align}\nonumber
H[N]&=\frac{\I}{2}\int_{\mathcal{N}}d^3v_o\,N(\bar{b}\dot{b}-\CC)+\int_{\mathcal{N}}d^3v_o\,p_\chi(\dot{N}+N\dot{\chi})+\\
&\quad+\frac{1}{2}\int_{\mathcal{N}}d^3v_o\,N\left(\pi_A\dot{\omega}^A-\dot{\pi}_A{\omega}^A-\utilde{\pi}_A\dot{\utilde{\omega}}^A+\dot{\utilde{\pi}}_A{\utilde{\omega}}^A\right).\label{H-cons2}
\end{align}

\paragraph{Gauss constraint} Consider now the oscillator modes $b$, $\bar{b}$ and the $U(1)$ generator $L=-\frac{1}{2}\mathrm{Tr}(\Pi J)=\frac{1}{2}J^{AB}\pi_A\omega_B$. On the kinematical phase space, they are independent. Physical configurations are constrained by \eref{b-def} and \eref{L-def}. This implies the constraint
\begin{equation}
\forall \lambda:\mathcal{N}\rightarrow\R\,\text{mod}\,2\pi:\lambda\big|_{\mathcal{C}_\pm}=0:G[\lambda]:=\int_{\mathcal{N}} d^3v_o\,\lambda\left(b\bar{b}+2Lr\right)=0.\label{G-cons}
\end{equation}
This constraint generates simultaneous $U(1)$ transformations of the oscillators $b$, $\bar{b}$ and the $SU(1,1)$ Heisenberg variables $\pi_A$, $\omega^A$.
\paragraph{Casimir-matching constraint} Finally, we have the \emph{matching conditions} which are a consequence of the fact that there is an $S\in SU(1,1)$ that maps the pair  $\pi^A$ and $\omega^A$ into $\utilde{\pi}^A$  and $\utilde{\omega}^A$, see \eref{t-piom-def}. This can happen if and only if the $SU(1,1)$ Casimirs $\pi_A\omega^A$ and $\utilde{\pi}_A\utilde{\omega}^A$ coincide. The $SU(1,1)$ Casimir can be real or imaginary. Thus, we distinguish two cases.
\begin{itemize}
\item \emph{Infra-Planckian modes, $L^2\geq c\bar{c}$.} In this case, $\pi_A\omega^A$ is imaginary. The matching condition for the $\mathfrak{su}(1,1)$ Casimir is 
\begin{equation}
\forall \lambda:\mathcal{N}\rightarrow\R,\lambda\big|_{\mathcal{C}_\pm}=0:M[\lambda]:=-\I\int_{\mathcal{N}} d^3v_o\,\lambda\left(\pi_A\omega^A-\utilde{\pi}_A\utilde{\omega}^A\right)=0.\label{M-cons1}
\end{equation}
This constraint generates simultaneous $U(1)$ rotations of $(\pi_A,\omega^A)$ and $(\utilde{\pi}_A,\utilde{\omega}^A)$, sending e.g.\ $(\pi_A,\omega^A)$ into $(\E^{\I\lambda}\pi_A,\E^{-\I\lambda}\omega^A)$.
\item \emph{Ultra-Planckian modes, $L^2\leq c\bar{c}$.} In this case, $\pi_A\omega^A$ is real. The matching condition for the $\mathfrak{su}(1,1)$ Casimir is 
\begin{equation}
\forall \lambda:\mathcal{N}\rightarrow\R,\lambda\big|_{\mathcal{C}_\pm}=0:M[\lambda]:=\int_{\mathcal{N}} d^3v_o\,\lambda\left(\pi_A\omega^A-\utilde{\pi}_A\utilde{\omega}^A\right)=0.\label{M-cons2}
\end{equation}
This constraint generates simultaneous dilatations of $(\pi_A,\omega^A)$ and $(\utilde{\pi}_A,\utilde{\omega}^A)$, sending e.g.\ $(\pi_A,\omega^A)$ into $(\E^{-\lambda}\pi_A,\E^{\lambda}\omega^A)$.\end{itemize}
\paragraph{Algebra of first-class constraints} Given the Poisson commutation relations on the kinematical phase space, i.e.\ \eref{b-oprts}, \eref{piom1}, \eref{piom2}, \eref{p-chi}, we obtain  the Poisson algebra of the first-class constraints
\begin{align}
\left\{H[N],H[N']\right\}&=-H[N\dot{N}'-\dot{N}N']\equiv H\left[[N,N']\right],\label{HH-brck}\\
\left\{H[N],G[\lambda]\right\}&=-G[N\dot{\lambda}],\label{HG-brck}\\
\left\{H[N],M[\lambda]\right\}&=-M[N\dot{\lambda}],\label{HM-brck}
\end{align}
where $N$ and $\lambda$ are appropriate smearing fucntions of compact support.
At the quantum level, the Poisson brackets become commutators with possibly anomalous terms on the right hand side. For example, \eref{HH-brck} turns into the Virasoro algeba with central charge $c_{\mtext{total}}$ that we compute in \hyperref[sec3]{Section 3}. At the classical level $G[\lambda]$ and $M[\lambda]$, which generate abelian symmetries, commute. At the quantum level, they are replaced by a $U(1)$ Kac--Moody algebra with central extension that vanishes as $\hbar\rightarrow 0$.
\subsection{Second-class constraints and their algebra}\label{sec2.5}
\noindent We introduced the Poisson brackets between $p_\xi$ and $\chi$ (chronoton modes), the oscillator modes $b$, $\bar{b}$ and the $SU(1,1)$ modes $(\pi_A,\omega^A;\utilde{\pi}_A,\utilde{\omega}^A)$ in  \eref{p-chi}, \eref{b-oprts}, \eref{piom1} and \eref{piom2}. This kinematical phase space has $2+2+4+4=12$ real dimensions. The relationship to the geometric data, namely the $U(1)$ connection $\Delta$, the shear $\sigma$ and the expansion $\vartheta$ is inferred from \eref{LieS}, \eref{Delt-def}, \eref{L-piom}, \eref{c-piom}, \eref{L-def}, \eref{c-def}. These relations impose constraints on the kinematical phase space. Three of these constraints are first-class. First of all, we have the $U(1)$ Gauss constraint \eref{G-cons}. In addition, there are two further first-class constraints, namely the Raychaudhuri condition \eref{H-cons1} and the matching constraints (\ref{M-cons1}, \ref{M-cons2}). The remaining four constraints, which we introduce in this section, are all second-class. The resulting physical phase space has $12-2\times 3-4=2$ local dimensions on phase space matching the two modes of gravitational radiation that can cross the null hypersurface.

\paragraph{Chronoton-matching constraint} First of all, we have to impose that the momentum $p_\chi$ of $\chi$ can be expressed in terms of the expansion of the null surface. Going back to \eref{theta-N1} and \eref{theta-N2}, we see that we need to match $p_\chi$ with the $U(1)$ generator, obtaining 
\begin{equation}
\forall f:\mathcal{N}\rightarrow\R,\lambda\big|_{\mathcal{C}_\pm}=0:\Upsilon[f]:=\int_{\mathcal{N}} d^3v_o\,f\left(p_\chi+2\gamma L\right)=0.\label{C-cons}
\end{equation}
At the quantum level, this constraint has the deparametrized form \cite{Giesel:2012,Giesel:2007wn,Husain:2011tk,Giesel:2016gxq,partobs,Domagala:2010bm} of a functional Schrödinger equation, in which $\chi$ plays the role of a multi-fingered clock and $L$, which is a $U(1)$ generator, is the Hamiltonian. It is also reminiscent of the diagonal simplicity constraint that plays an important role in the definition of the EPRL spinfoam amplitudes \cite{spezialetwist2,LQGvertexfinite,PhysRevD.82.064026,flppdspinfoam,Wieland:2013cr}. However, there is also a difference. The constraint  forms a second class system  together with the rest of the constraint algebra. At the quantum level, the constraint can only be imposed weakly, e.g.\ by restricting the smearing functions  to a superposition of only positive frequency modes, i.e.\ $f\sim\sum_{n> 0} f_n\E^{-\I n u}$. It would be also interesting to investigate a gauge unfixing approach, in which \eref{C-cons} is promoted into a first-class constraint at the expense of treating one of the second-class constraints\footnote{The relevant constraint is the $U(1)$ angle-matching constraint $\Phi[f]$ introduced below, see \eref{Phi-cons} and \eref{Up-Phi-brck}.} as a mere gauge-fixing condition, see \cite{Vytheeswaran:1994np}.
\paragraph{Components of the Maurer--Cartan form} Finally, we have three more constraints that impose  constraints between the time component of the Maurer--Cartan form $\di S S^{-1}$ and the canonical variables $b$, $\bar{b}$ and $(\pi_A,\omega^A;\utilde{\pi}_A,\utilde{\omega}^A)$.

Before discussing the details, let us consider first the $\mathfrak{su}(1,1)$ element $\mu=\dot{S}S^{-1}$, see \eref{LieS} and express it in terms of the $SU(1,1)$ Heisenberg variables $(\pi_A,\omega^A;\utilde{\pi}_A,\utilde{\omega}^A)$. Going back to \eref{t-piom-def}, we notice first
\begin{equation}
\ou{h}{A}{B}:=\omega^A\utilde{\pi}_B-\pi^A\utilde{\omega}_B=\utilde{\pi}_C\utilde{\omega}^C \ou{S}{A}{B}.\label{h-def}
\end{equation}
Notice that the bilinear observable $\ou{h}{A}{B}$ Poisson commutes with the matching constraints (\ref{M-cons1},\ref{M-cons2}). Moving forward, we restrict ourselves to the $M=0$ constraint hypersuface.
Equation \eref{h-def}  implies that on this hypersuface the trace free part of $\ou{\dot{h}}{A}{C}\uo{h}{B}{C}=\epsilon_{BE}\epsilon^{CF}\ou{\dot{h}}{A}{C}\ou{h}{E}{F}$ is determined by
\begin{equation}
{\dot{h}}{}_{(A}{}^{C}h_{B)C}=j\,\utilde{j}\,{\dot{S}}{}_{(A}{}^{C}S_{B)C},
\end{equation}
where we introduced the abelian currents
\begin{equation}
j=\pi_A\omega^A,\qquad \utilde{j}=\utilde{\pi}_A\utilde{\omega}^A.
\end{equation}
The two missing constraints can be inferred from \eref{c-piom}, \eref{Delt-def}, \eref{c-def}, \eref{LieS}. These two constraints impose that $\dot{S}S^{-1}\in \mathfrak{su}(1,1)$ determines the $U(1)$ angle $\varphi=-\I(\dot{b}\bar{b}-\CC)\propto\mathrm{Tr}(J\Pi)$ and the shear $\sigma\propto\mathrm{Tr}(\Pi X)$ (one complex constraint).   
\paragraph{Angle-matching constraint} The diagonal $J$-component of $\dot{S}S^{-1}$ is given by the $U(1)$ spin connection coefficient $\varphi$, see \eref{Lie-m}, \eref{LieS}. The corresponding $U(1)$ holonomy $\E^{\I \Delta}$ enters the definition of the $b$, $\bar{b}$ oscillator modes in terms of the geometric data, see \eref{Delt-def}, \eref{b-def}. At the kinematical level, where $b$, $\bar{b}$ and the $SU(1,1)$ modes $(\pi_A,\omega^A;\utilde{\pi}_A,\utilde{\omega}^A)$ are independent, these conditions may be violated. To impose them among the kinematical variables, we  need to impose the constraint
\begin{equation}
\begin{split}
&\forall f:\mathcal{N}\rightarrow\R,f\big|_{\mathcal{C}_\pm}=0:\\
&\Phi[f]:=\frac{1}{2}\int_{\mathcal{N}} d^3v_o\,f\left(b\bar{b}\,\uo{\dot{h}}{(A}{C}h_{B)C}J^{AB}+\I\,r \,j\,\utilde{j}\,\big(\dot{b}\bar{b}-\CC\big)\right)=0.\label{Phi-cons}
\end{split}
\end{equation}
Notice that this is a polynomial of degree six on the kinematical phase space. All other constraints that we mentioned so far are at most quadratic, see \eref{H-cons2}, \eref{G-cons}, (\ref{M-cons1}, \ref{M-cons2}), \eref{C-cons}.  At the quantum level, we will need to introduce (conformal) ordering, but this is not difficult to handle. In here, only a few contractions appear. For example, the normal ordering of the second term in \eref{Phi-cons} is simply $\normord{j\utilde{j}\dot{b}\bar{b}}=\normord{\pi_A\omega^A}\,\normord{\utilde{\pi}_B\utilde{\omega}^B}\,\normord{\dot{b}\bar{b}}$.
\paragraph{Shear-matching constraint} The off-diagonal $X$-component of $\dot{S}S^{-1}$ is given by the shear $\sigma$ of the null generators, see \eref{Lie-m}, \eref{LieS}. The shear $\sigma$ times the conformal factor $\Omega^2$ enters the definition of the $\mathfrak{su}(1,1)$-valued momentum variable $\ou{\Pi}{A}{B}=\frac{1}{2}(\pi^A\omega_B+\omega^A\pi_B)$, see \eref{Pi-def}, \eref{c-def}, \eref{c-piom}. At the kinematical level, where $b$, $\bar{b}$ and the $SU(1,1)$ modes $(\pi_A,\omega^A;\utilde{\pi}_A,\utilde{\omega}^A)$ are independent, these conditions may be violated. To impose them among the kinematical variables, we need to impose the constraint
\begin{equation}
\begin{split}
&\forall f:\mathcal{N}\rightarrow\R,f\big|_{\mathcal{C}_\pm}=0:\\
&\Xi[f]:=\int_{\mathcal{N}} d^3v_o\,f\,X^{AB}\left(j\,\utilde{j}\,\pi_{(A}\omega_{B)}+\frac{1}{8\pi\gamma G}(\gamma+\I)\Omega^2\uo{\dot{h}}{(A}{C}h_{B)C}\right)=0.\label{Xi-cons}
\end{split}
\end{equation}
In here, $\Omega^2$ is the conformal factor, which can be obtained as a kinematical observable on phase space by integrating the defining equations of the oscillator modes. Going back to \eref{a-def} and \eref{b-def}, we obtain $\Omega^2$ by solving
\begin{equation}
\frac{\di}{\di u}\Omega^2=8\pi\gamma G\,r\, b\bar{b},\qquad\Omega^2\big|_{u=2\pi}=8\pi\gamma G\, a\bar{a}.
\end{equation}

\paragraph{Algebra of second-class constraints} The computation of the constraint algebra involving the second-class constraints is straight-forward but not very illuminating. Here, we only give the final result. For brevity, we also drop all terms that vanish on the constraint hypersurface. Moving ahead, the symbol \qq{$\approx$} means equality up to terms that vanish as we set $H=0, G=0, M=0$ and $\Upsilon=0, \Phi=0, \Xi=0$. On that hypersurface in phase space, the only non-vanishing Poisson brackets among the constraints are
\begin{align}
\left\{\Upsilon[f],\Phi[f']\right\}&\approx-\gamma\int_{\mathcal{N}}d^3v_o\,f'\,\dot{f}\,b\bar{b}\,j\,\utilde{j},\label{Up-Phi-brck}\\
\left\{\Phi[f],\Xi[f']\right\}&\approx-\gamma\int_{\mathcal{N}}d^3v_o\,f\,f'\, j^2\,\utilde{j}^2\,b\bar{b}\,\sigma,\label{Phi-Xi-brck}\\
\left\{\Xi[f],\bar{\Xi}[\bar{f}']\right\}&\approx-\frac{1}{8\pi G}\int_{\mathcal{N}}d^3v_o\left(f\E^{2\I\Delta}\frac{\di}{\di u}\left(\bar{f}'\E^{-2\I\Delta}\right)-\bar{f}'\E^{-2\I\Delta}\frac{\di}{\di u}\left(f\E^{2\I\Delta}\right)\right)\Omega^2\,j^2\,\utilde{j}^2.\label{Xi-bXi-brck}
\end{align}
These three lines determine the Dirac matrix and thereby the resulting Dirac bracket on the constraint hypersurface. We do not give the explicit construction here, see \cite{Wieland:2021vef} for further details, where a similar analysis was done using directly the $T^\ast SL(2,\R)$ algebra.

\section{Quasi-local Fock representation}\label{sec3}
\subsection{Mode expansion, reality conditions and Fock vacuum}
\noindent We now return to our main problem. The task ahead is to find a Hilbert space representation of the boundary modes and investigate the spectrum of the radiated power. The first step is to consider a Fourier \emph{series} expansion of the boundary fields and separate them into positive and negative frequency modes. That there is a Fourier series rather than a Fourier integral is the result of an implicit IR cutoff.\footnote{This cutoff is different from a Lorentz violating large distance cutoff of QED, in which we put all fields into a box of finite physical volume $L^3=\int_{\mathcal{B}}\di^3x$ and impose reflecting boundary conditions at $\partial{\mathcal{B}}$. In our case, the data that determines the physical extension of the null surface $\mathcal{N}$ has a quantum description as well. The size of $\mathcal{N}$ is subject to quantum fluctuations. This is different from an IR cutoff, in which $L$, which measures the size of the box, is a mere $c$-number. Take, for example, the difference between the initial and final area of the two cuts. This is simply the global number operator on Fock space counting the quantum excitations of the $b$, $\bar{b}$ modes.} The null surface $\mathcal{N}$ has an initial and final spacelike boundary $\mathcal{C}_\pm$, where the auxiliary null coordinate $u$ assumes boundary values $u=0$ and $2\pi$ respectively. The positive frequency modes annihilate the vacuum, negative frequency modes excite quasi-local graviton modes. The underlying Fock vacuum is akin to the pre-geometric (no geometry) Ashtekar--Lewandowski vacuum of loop quantum gravity \cite{LOSTtheorem,Ashtekar:1994mh}. 
\paragraph{Chronoton modes} The mode expansion for the chronoton modes is
\begin{align}
\chi=\frac{1}{\sqrt{2\pi}}\sum_{n=-\infty}^\infty \chi_n(\zeta,\bar{\zeta})\E^{-\I n u},\\
p_\chi=\frac{1}{\sqrt{2\pi}}\sum_{n=-\infty}^\infty p_{\chi n}(\zeta,\bar{\zeta})\E^{-\I n u}.
\end{align}
The kinematical Poisson brackets \eref{p-chi} imply 
\begin{equation}
\left\{p_{\chi n}(\zeta,\bar{\zeta}),\chi_{m}(\zeta',\bar{\zeta}')\right\}=\delta_{m+n}\delta_{\mathcal{C}}(\zeta-\zeta',\bar{\zeta}-\bar{\zeta}').\label{p-chi-mods-pcmmt}
\end{equation}
Since $p_\chi$ and $\chi$ are real, see \eref{chi-def} and \eref{theta-N1}, from which we inferred that $p_\chi=(8\pi G)^{-1}\frac{\di}{\di u}\Omega^2$, we obtain the reality conditions
\begin{equation}
\overline{\chi_n}=\chi_{-n},\qquad\overline{{p}_{\chi n}}=p_{\chi|-n}.\label{chi-p-realcond}
\end{equation}

\paragraph{Area modes} In the same way, we introduce the mode expansion for the area modes, which are complex-valued
\begin{align}
b=\frac{1}{\sqrt{2\pi}}\sum_{n=-\infty}^\infty b_n(\zeta,\bar{\zeta})\E^{-\I n u},\\
\bar{b}=\frac{1}{\sqrt{2\pi}}\sum_{n=-\infty}^\infty \bar{b}_n(\zeta,\bar{\zeta})\E^{-\I n u}.
\end{align}
The kinematical Poisson brackets \eref{p-chi} imply 
\begin{equation}
\left\{b_{n}(\zeta,\bar{\zeta}),\bar{b}_{m}(\zeta',\bar{\zeta}')\right\}=\I\,\delta_{m+n}\delta_{\mathcal{C}}(\zeta-\zeta',\bar{\zeta}-\bar{\zeta}').\label{bb-mods-pcmmt}
\end{equation}
The reality conditions are
\begin{equation}
\overline{b_{n}}=\bar{b}_{-n}.\label{realcond-b}
\end{equation}

\paragraph{$SU(1,1)$ modes} Finally, there are the $SU(1,1)$ modes, which are complex. We introduce the mode expansion
\begin{align}
\pi_A=\frac{1}{\sqrt{2\pi}}\sum_{n=-\infty}^\infty \pi_{An}(\zeta,\bar{\zeta})\E^{-\I n u},\\
\omega^A=\frac{1}{\sqrt{2\pi}}\sum_{n=-\infty}^\infty \ou{\omega}{A}{n}(\zeta,\bar{\zeta})\E^{-\I n u}.
\end{align}
and equally
\begin{align}
\utilde{\pi}_A=\frac{1}{\sqrt{2\pi}}\sum_{n=-\infty}^\infty \utilde{\pi}_{An}(\zeta,\bar{\zeta})\E^{-\I n u},\\
\utilde{\omega}^A=\frac{1}{\sqrt{2\pi}}\sum_{n=-\infty}^\infty \ou{\utilde{\omega}}{A}{n}(\zeta,\bar{\zeta})\E^{-\I n u}.
\end{align}
The Poisson brackets on the kinematical phase space \eref{piom1} and \eref{piom2} translate into
\begin{align}
\left\{\pi_{An}(\zeta,\bar{\zeta}),\omega_{Bm}(\zeta',\bar{\zeta}')\right\}&=+\epsilon_{AB}\delta_{n+m}\delta_{\mathcal{C}}(\zeta-\zeta',\bar{\zeta}-\bar{\zeta}'),\label{pi-om-mods-pcmmt1}\\
\left\{\utilde{\pi}_{An}(\zeta,\bar{\zeta}),\utilde{\omega}_{Bm}(\zeta',\bar{\zeta}')\right\}&=-\epsilon_{AB}\delta_{n+m}\delta_{\mathcal{C}}(\zeta-\zeta',\bar{\zeta}-\bar{\zeta}'),\label{pi-om-mods-pcmmt2}
\end{align}
These modes are complex and they need to satisfy reality conditions \eref{realcond1} and \eref{realcond2}. Depending on the value of the  $SU(1,1)$ Casimir, two cases need to be distinguished. Assuming the fundamental fields are bosonic, we obtain\footnote{For the case in which the $SU(1,1)$ modes are anti-commuting Grassmann-valued, see \hyperref[sec3.3]{Section 3.3} below.}
\begin{itemize}
\item \emph{Infra-Planckian modes, $L^2\geq c\bar{c}$.} In this case, $\pi_A\omega^A$ is imaginary. The reality conditions for the $SU(1,1)$ modes are
\begin{equation}
\pi_{An}=\pm\I\,\eta_{A\bar B}\,\overline{\ou{\omega}{B}{-n}},\qquad\omega_{An}=\pm\I\,\eta_{A\bar{B}}\,\overline{\ou{\pi}{B}{-n}}.\label{modecond1}
\end{equation}
\item \emph{Ultra-Planckian modes, $L^2< c\bar{c}$.} In this case, $\pi_A\omega^A$ is real. The reality conditions for the $SU(1,1)$ modes are
\begin{equation}
{\pi}_{An}=\eta_{A\bar B}\,\overline{\ou{\pi}{B}{-n}},\qquad{\omega}_{An}=\eta_{A\bar{B}}\,\overline{\ou{\omega}{B}{-n}}.\label{modecond2}
\end{equation}
Since the two sectors are related by an $SU(1,1)$ transformation that preserves the $\eta_{A\bar{A}}$-inner product, see \eref{SU1-1-invrncs}, \eref{t-piom-def}, the same conditions also apply to $\utilde{\pi}_{An}$ and $\ou{\utilde{\omega}}{A}{n}$.
\end{itemize}
A good intuition for these two different cases can be found by considering the follwing simplified model. Consider the complex phase space $\C^2\ni(p,z)$ space with Poisson brackets $\{p,z\}=1$. The \emph{infra-Planckian modes} correspond to the reality conditions $p=\pm\I\bar{z}$, turning $z$ and $\bar{z}$ into ordinary (bosonic or fermionic) creation and annhilation operators, such that e.g.\ $a=1/\sqrt{\hbar}\,z$. A basis in the Hilbert space is given by the ordinary oscillator states $|n\rangle=(n!)^{-1/2}\,a^n|0\rangle$. The \emph{ultra-Planckian modes} behave differently. They satisfy reality conditions $p=\bar{p}$ and $z=\bar{z}$. The resulting physical Hilbert space is simply $L^2(\R,\di x)\ni\Psi$, operators act as $\langle x|z\Psi\rangle=x\Psi(x)$,  $\langle x|p\Psi\rangle=-\I\hbar\partial_x\Psi(x)$. In short, the choice of reality conditions determines the choice of Hilbert space inner product \cite{ashtekar}.

\paragraph{Quasi-local Fock vacuum} To quantise these modes, we consider $N$ distinct points (punctures) at angular coordiantes $(\zeta_i,\bar{\zeta}_i)$, $i=1\dots N$. Next, we introduce a dual tessellation of the cross-sectional geometry, $\mathcal{C}=\bigcup_{i=1}^N\mathcal{D}_i$, such that each puncture lies in only one plaquette. Consider then the chronoton $\chi$ and its momentum variable $p_\chi$. As far as the two-dimensional geometry is concerned, $\chi$ is a scalar and $p_\chi$ is a density of weight one. We can thus naturally smear the momentum over the $i$-th plaquette obtaining
\begin{equation}
p_{\chi n}(i):=\int_{\mathcal{D}_i}d^2v_o\,p_{\chi n}(\zeta,\bar{\zeta})
\end{equation}
In addition, we set
\begin{equation}
\chi_{n}(i):=\chi_{n}(\zeta_i,\bar{\zeta}_i).
\end{equation}
The Poisson brackets \eref{p-chi-mods-pcmmt} imply
\begin{equation}
\left\{p_{\chi n}(i),\chi_{m}(j)\right\}=\delta_{m+n}\delta_{i,j},\qquad i=1,\dots,N.\label{pchi-poiss}
\end{equation}
In the same way, we proceed with the remaining variables. The area oscillators $b_n(\zeta,\bar{\zeta})$, $\bar{b}_n(\zeta,\bar{\zeta})$ are half-densities \cite{ashtekar, thiemann, Thiemann:1997rq} on the two-dimensional cut. Following \cite{Thiemann:1997rq,Wieland:2017cmf}, we remove the density half-weights by integrating the oscillator modes against the square root of the two-dimsnsional Dirac delta distribution obtaining
\begin{equation}
b_n(i)=\int_{\mathcal{D}_i}d^2v_o(\zeta,\bar{\zeta})\,\sqrt{\delta_{\mathcal{C}}(\zeta-\zeta_i,\bar{\zeta}-\bar{\zeta}_i)}\,b_n(\zeta,\bar{\zeta}),
\end{equation}
and equally for $\bar{b}_n(\zeta,\bar{\zeta})$ and the edge-mode oscillators $a(\zeta,\bar{\zeta})$ and $\bar{a}(\zeta,\bar{\zeta})$. The resulting regularized Poisson brackets are
\begin{align}
\left\{b_n(i),\bar{b}_m(j)\right\}&=\I\,\delta_{m+n}\delta_{i,j},\quad i,j=1,\dots N,\\
\left\{a(i),\bar{a}(j)\right\}&=\I\,\delta_{i,j},\quad i,j=1,\dots N.
\end{align}
Repeating the construction for the  $SU(1,1)$ modes, we obtain
\begin{align}
\left\{\pi_{An}(i),\omega_{Bm}(j)\right\}&=+\epsilon_{AB}\delta_{n+m}\delta_{i,j},\quad i,j=1,\dots,N,\label{pi-om-pnct1}\\
\left\{\utilde{\pi}_{An}(i),\utilde{\omega}_{Bm}(j)\right\}&=-\epsilon_{AB}\delta_{n+m}\delta_{i,j},\quad i,j=1,\dots,N\label{t-pi-om-pnct2}.
\end{align}

\paragraph{Fock vacuum} It is now immediate to pass ahead to quantum theory, where Poisson brackets turn into (anti)-commutators $\frac{\I}{\hbar}[\cdot,\cdot]_\pm$, $[A,B]_\pm =AB\mp BA$, depending on the choice of underlying Bose (Fermi) statistics. The correct statistics will be identified in the next section. Before moving ahead, a few final remarks for this section.  The  Fock vacuum $|0\rangle$ is now readily identified as the state annihilated by the area edge mode $a(i)$
\begin{equation}
a(i)|0\rangle=0, \quad \forall i=1,\dots,N.\label{Fock-vac1}
\end{equation}
and all positive frequency modes, i.e.\
\begin{equation}
x_n(i)|0\rangle=0, \quad \forall n>0,\quad i=1,\dots,N.\label{Fock-vac2}
\end{equation}
where $x_n$ can stand for all modes, i.e.\ the area modes $b_n,\bar{b}_n$, the chronoton modes $p_{\chi n},\chi_n$,  and the $SU(1,1)$ modes $\pi_{An},\ou{\omega}{A}{n},\utilde{\pi}_{An},\ou{\utilde{\omega}}{A}{n}$ that encode the radiative data (shear). The inner product is then inferred purely algebraically from \eref{Fock-vac1} and \eref{Fock-vac2} and the canonical (anti)-commutation relations. However, there is a problem. The $SU(1,1)$ inner product is not positive-definite. There are negative norm states. This implies that the resulting (kinematical) Hilbert space will have negative norm states as well. This is in complete analogy with the tachyonic modes that we have in string theory \cite{polchinski,green}. In string theory, they arise from the signature $(-$$+$$+$$+\dots)$ Lorentzian inner product of the background spacetime. In here, they arise on the Wheelerian {super-space} \cite{Misner:1973prb,Giulini:2009np} of the null surface $\mathcal{N}$, i.e.\ the space  of null initial data parametrized by the choice of canonical variables.\footnote{{N.B.:} on a spacelike hypersurface, the Wheeler--De Witt supermetric is not positive-definite either \cite{DeWitt:1967yk}.} We could, of course, say to define the Fock vacuum differently. Let us briefly mention this possibility. Choose a dyad $(o^A,\iota^A)$ in $\C^2$ such that $\eta_{A\bar{A}}o^A\bar{o}^{\bar{A}}=-1$ and $\eta_{A\bar{A}}\iota^A\bar{\iota}^{\bar{A}}=+1$ and define the components on superspace $\alpha_n=\omega_{An}\iota^A$ and $\beta_n^\dagger=\omega_{An}o^A$. Consider, for concreteness the infra-Planckian modes \eref{modecond1}. It is then always possible to define an inner product that is positive-definite on the entire state space by demanding that $\alpha_n$ and $\beta_n$ annihilate the vacuum \emph{for all} $n\in\Z$. The problem with this construction is that it is at odds with the conformal field theory methods (radial ordering, state-operator correspondence) that we employ in the next section to construct representations of the Virasoro algebra \eref{HH-brck} on $\mathcal{N}$ with non-trivial central charge. Furthermore, the resulting vacuum state would no longer be invariant under global $SU(1,1)$ rotations, thus selecting a preferred frame. Thus, we no longer follow this route.

\paragraph{Further remarks: Holding $N$ fixed, lack of entanglement} In what follows, we keep $N$ fixed. Thus, for each $N\geq 0$, we obtain a different Hilbert space $\mathcal{H}_N$, with $\mathcal{H}_{N=0}=\C$. It seems logical to allow for arbitrary high particle number $N$, thus working on an extended Hilbert space $\mathcal{F}=\bigoplus_{N=0}^\infty\mathcal{H}_N$. Such a Hilbert space would arise naturally from a second quantisation of the present first-quantised framework. Thus, the one-particle wave function, i.e.\ $\Psi\in\mathcal{H}_{N=1}$, would turn into a quantum field on the superspace of a single null generator. We will not investigate this possibility here further, but it clearly resonates with the group field theory approach to quantum gravity \cite{oriti,Freidel:2005qe,Oriti:2013aqa,Gielen:2024sxs}. Another, but possibly related viewpoint, is to work at fixed $N$ before then taking an inductive $N\rightarrow\infty$ limit. This would require a suitable choice of embedding and refinement maps, see e.g.\ \cite{Rovelli:2010qx,Bahr:2012qj,Dittrich:2014ala}. Either viewpoint may be useful to address, in fact, another limitation of the current approach: generic states in $\mathcal{F}$ will have an entanglement structure, which is in conflict with a smooth classical geometry on each cut. We expect that this problem can be resolved by selecting a subspace of $\mathcal{F}$ in which the (super)-angular momentum charges $J_\xi$, or rather their finite generators $\exp(-\frac{\I}{\hbar} J_\xi)$ with $\xi^a\in T\mathcal{C}$, have the correct short-distance entanglement. A similar viewpoint has been advocated in \cite{Cao:2016mst}, where the spatial geometry is reconstructed from the entanglement structure of the Hamiltonian.

\subsection{Auxiliary conformal field theory}
\noindent In the previous section, we identified the classical constraint algebra, see \eref{HH-brck}, \eref{HG-brck}, \eref{HM-brck} and \eref{Up-Phi-brck}, \eref{Phi-Xi-brck}, \eref{Xi-bXi-brck}. To obtain a representation of this algebra at the quantum level, we consider first a few basic fact about two-dimensional conformal field theory (CFT) and the $\beta\gamma$ ($bc$) system \cite{CFTbook}.

\paragraph{Preparatory remarks} We introduce the auxiliary euclidean action
\begin{equation}
S[p,q]=\frac{1}{\I\hbar}\int_{\C} \di^2 z\,p_i\bar{\partial}q^i,\qquad \di^2z=\I\,\di z\wedge\di\bar{z},\label{aux-actn}
\end{equation}
where $z=x+\I y$ and $\partial=\tfrac{1}{2}(\partial_x+\I \partial_y)$. Indices of the configuration and momentum space fields $p_i$ and $q^i$ refer to some possibly complex $d$-dimensional vector space. In the following, we will leave the underlying statistic unspecified. Accordingly, $[A,B]_\pm=AB \mp BA$ denotes the (anti)-commutator. A trivial field redefinition brings this action into the standard form of the $\beta\gamma$-system ($bc$-system) of superstring theory.\smallskip

The euclidean path integral for the auxiliary CFT \eref{aux-actn} determines the $n$-point functions
\begin{equation}
\left\langle\mathcal{R}\left(\mathcal{O}_1(z_1,\bar{z_1})\mathcal{O}_2(z_2,\bar{z_2})\dots\right)\right\rangle=\int\mathcal{D}p\,\mathcal{D}q\,\mathcal{O}_1(z_1,\bar{z_1})\mathcal{O}_2(z_2,\bar{z_2})\dots\E^{-S[p,q]},\label{n-pntfnctns}
\end{equation}
where $\mathcal{R(\dots)}$ denotes radial ordering,
\begin{equation}
\mathcal{R}\left(\mathcal{O}_1(z_1,\bar{z_1})\mathcal{O}_2(z_2,\bar{z_2})\right)=\begin{cases} 
\phantom{\pm}\mathcal{O}_1(z_1,\bar{z_1})\mathcal{O}_2(z_2,\bar{z_2}),\qquad |z_2|<|z_1|\\
\pm\mathcal{O}_2(z_2,\bar{z_2})\mathcal{O}_1(z_1,\bar{z_1}),\qquad |z_1|<|z_2|.
\end{cases}
\end{equation}
Time evolution is replaced by radial evolution. The $n$-point functions can be understood as expectation values between an asymptotic in-vacuum at asymptotic Euclidean initial time $|z|\rightarrow 0$ and an out-vacuum at $|z|\rightarrow \infty$ that are sandwiched between an Euclidean radial evolution operator. 
 The field operators $p_i$ and $q^i$ satisfy the Heisenberg operator equations 
\begin{equation}
\bar\partial q^i=0,\qquad\bar\partial p_i=0,\label{field-eqs}
\end{equation}
hence $p_i$ and $q^i$ are holomorphic. Thus the fields are necessarily complex. Hermitian conjugation ($\mathcal{O}\rightarrow\mathcal{O}^\dagger)$ is realized as a combination of time reversal $(z\mapsto \bar{z}^{-1})$ and charge conjugation $((p_i,q^i)\rightarrow(\bar{p}_i,\bar{q}^i))$, 
\begin{equation}
[p_i]^\dagger(z)=\overline{p_i(\bar{z}^{-1})},\qquad [q^i]^\dagger(z)=\overline{q^i(\bar{z}^{-1})}.
\end{equation}
Finally, we have the conformal normal ordering \qq{$\normord{\cdots}$} of operator products, which is defined for bosonic (fermionic) fields through Wick's theorem and
\begin{equation}
\mathcal{R}\left(p_i(z_1)q^j(z_2)\right)=\normord{p_i(z_1)q^j(z_2)}\mp\frac{\I\hbar}{2\pi}\frac{1}{z_{12}}\delta^j_i,\label{normord-def}
\end{equation}
where $z_{12}=z_1-z_2$.

Next, we introduce the mode expansion
\begin{align}
p_i(z)&=\frac{1}{\sqrt{2\pi}}\sum_{n=-\infty}^\infty p_{in}\frac{1}{z^{n+\lambda}},\\
q^i(z)&=\frac{1}{\sqrt{2\pi}}\sum_{n=-\infty}^\infty \ou{q}{i}{n}\frac{1}{z^{n-\lambda+1}}.
\end{align}
This can be inverted immidately to give
\begin{align}
p_{in}&=\frac{1}{\sqrt{2\pi}\,\I}\oint\di z\,z^{n+\lambda-1}p_i(z),\\
\ou{q}{i}{n}&=\frac{1}{\sqrt{2\pi}\,\I}\oint\di z\,z^{n-\lambda}q^i(z),
\end{align}
where the  path for the integration encircles the origin in counterclockwise direction. Equation \eref{normord-def} implies now immediately the Heisenberg (anti)-commutation relations
\begin{equation}
[p_{in},\ou{q}{j}{m}]_{\pm}=p_{in}\ou{q}{j}{m}\mp \ou{q}{j}{m}p_{in}=\mp\I\hbar\delta_{m+n}.\label{H-berg-rel}
\end{equation}
\paragraph{CFT auxiliary stress-energy tensor} We define the auxiliary stress energy tensor,
\begin{equation}
T(z)=\frac{2\pi}{\I\hbar}\Big[(1-\lambda)\normord{(\partial p_i)(z)q^i(z)}-\lambda\normord{p_i(z)(\partial q^i)(z)}+v^i\partial p_i(z)\Big],\label{T-def}
\end{equation}
where $v^i:\partial v^i=0=\bar{\partial}v^i$ is a fixed target-space vector. The corresponding Ward identities are
\begin{equation}
\delta_\xi\mathcal{O}=\frac{1}{2\pi \I}\oint\di z\,\xi(z)\,\mathcal{R}\big(T(z)\mathcal{O}\big),\label{Ward-ident}
\end{equation}
where $\delta_\xi$ is the quantisation of the classical Lie derivative $\mathcal{L}_\xi$, which is taken here with respect to the conformal Killing field $\xi^a=\xi(z)\partial^a_z+\bar{\xi}(\bar{z})\partial^a_{\bar{z}}$. The (anti)-commutation relations now immediately imply
\begin{align}
\delta_\xi p_i(z)&=\xi(z)\left(\partial p_i\right)(z)+\lambda\left(\partial\xi\right)(z)\,p_i(z),\\
\delta_\xi q^i(z)&=\xi(z)\left(\partial q^i\right)(z)+(1-\lambda)\left(\partial\xi\right)(z)\,q^i(z)\pm v^i(\partial\xi)(z).
\end{align}
From the operator product expansion of the stress energy tensor with itself, we infer the central charge. For bosonic (fermionic) commutation relations, we obtain
\begin{equation}
T(z)T(0)\sim \frac{1}{z}(\partial T)(0)+\frac{2}{z^2}T(0)\pm\frac{1}{z^4}d(6\lambda^2-6\lambda+1),
\end{equation}
where \qq{$\sim$} means equality up to all non-singular terms as $z\rightarrow 0$. Thus, the central charge is
\begin{equation}
c=\begin{cases}+d\left(3(2\lambda-1)^2-1\right)\quad\text{if $p_i$ and $q^i$ are bosons,}\\
-d\left(3(2\lambda-1)^2-1\right)\quad \text{if $p_i$ and $q^i$ are fermions.}\end{cases}
\end{equation}

\paragraph{Abelian current} Another important observable is the abelian current,
\begin{equation}
j(z)=\normord{p_i(z)q^i(z)}
\end{equation}
The corresponding Ward identity is
\begin{equation}
\delta_\lambda^{U(1)}\mathcal{O}=\frac{1}{\hbar}\oint\di z\,\lambda(z)\,\mathcal{R}\left(j(z)\mathcal{O}\right).
\end{equation}
The (anti)-commutation relations \eref{normord-def} imply the abelian symmetry
\begin{align}
\delta_\lambda^{U(1)}q^i(z)&=+\lambda(z)q^i(z),\\
\delta_\lambda^{U(1)}p_i(z)&=-\lambda(z)p_i(z).
\end{align}
Upon imposing reality conditions for $p_i$ and $q^i$, e.g.\ \eref{modecond1} or \eref{modecond2}, we need to impose also conditions on the gauge parameter $\lambda$ as well, e.g.\ $\bar{\lambda}(\bar{z}^{-1})=\varepsilon\,\lambda(z)$, where $\varepsilon\in\{-1,1\}$.\smallskip

The next step ahead is to match these common definitions \cite{CFTbook} with the boundary fields and their transformations introduced above, see \eref{Lxi-b}, \eref{Lxi-bbar}, \eref{Lxi-chi}, \eref{Lxi-pchi}, \eref{Lxi-piom}, \eref{Lxi-t-piom}.

\paragraph{Auxiliary $CFT$ for the chronoton modes} Classical Poisson brackets $\{A,B\}$ between observables $A$ and $B$ arise from the semi-classical limit of $\frac{\I}{\hbar}$ times the (anti)-commutator $[A,B]_\pm$. To match the classical Poisson commutation relations \eref{pchi-poiss} with the Heisenberg relations \eref{H-berg-rel}, we can thus identify $(p_i,q^i)$ with $(p_\chi,\chi)$. The action of the Hamiltonian constraint that generates the time reparametrizations of the null coordinate has been given in \eref{Lxi-chi} and \eref{Lxi-pchi} aove. Comparison with \eref{T-def} allows us to introduce the corresponding stress energy tensor for the chronoton modes
\begin{equation}
T^{(\chi)}(z)=\frac{2\pi\I}{\hbar}\Big[\normord{p_\chi(z)(\partial\chi)(z)}\pm(\partial p_\chi)(z)\Big],
\end{equation}
thus $\lambda=1$. The central charge for the chronoton system on each plaquette is therefore given by
\begin{equation}
c^{(\chi)}=\pm2,
\end{equation}
depending on wether the chronotons are bosons and fermions. Consider now the reality conditions \eref{chi-p-realcond}. The Heisenberg relations \eref{H-berg-rel} immediately imply
\begin{equation}
[p_{\chi n},\chi_m]_{\pm}=[p_{\chi|-n},\chi_{-m}]_{\pm}.
\end{equation}
Hermitian conjugation flips the order of the (anti)-commutator, i.e.\ $[A,B]^\dagger_\pm=[B^\dagger,A^\dagger]_\pm=\mp[A^\dagger,B^\dagger]$. This implies that only the bosonic commutation relations are consitsent with the reality conditions \eref{chi-p-realcond}. Therefore, the central charge for the chronoton system is
\begin{equation}
c^{(\chi)}=2.
\end{equation}
\paragraph{Auxiliary $CFT$ for the area modes} We consider the classical Poisson brackets \eref{bb-mods-pcmmt} and compare them with the Heisenberg relations of the auxiliary CFT, see \eref{H-berg-rel}. This allows us to make the following identifications.  The ${b}_n$-modes correspond to $\ou{q}{i}{n}$ and the $\bar{b}_{n}$ modes correspond to $\I\, p_{in}$. We can then also identify the auxiliary stress energy tensor of the $b_n$-modes as 
\begin{equation}
T^{(b)}=\frac{\pi}{\hbar}\Big[\normord{(\partial\bar{b})(z)b(z)}-\normord{\bar{b}(z)(\partial b)(z)}\Big].
\end{equation}
The central charge is
\begin{equation}
c^{(b)}=\mp1,
\end{equation}
depending on whether we choose bosonic ($-$) or fermionic ($+$) commutation relations. For either choice, the Heisenberg relations \eref{H-berg-rel} are
\begin{equation}
\left[b_n,\bar{b}_{m}\right]_\pm=\hbar\,\delta_{m+n}.
\end{equation}
Notice that the reality conditions \eref{realcond-b}, i.e.\ $b^\dagger_n=b_{-n}$, are consistent with both bosonic and fermionic commutation relations.

\paragraph{Auxiliary $CFT$ for the $SU(1,1)$ modes} We are now left with the $SU(1,1)$ modes. For the moment, let us only consider the bosonic case. We deal with the fermionic case, where the reality conditions \eref{realcond1} and \eref{realcond2} are slightly modified, separately in \hyperref[sec3.3]{Section 3.3} below. Consider then the classical Poisson brackets on each plaquette, which are given in \eref{pi-om-pnct1} and \eref{t-pi-om-pnct2}. As in above, we can thus identify $(\pi_{An},\utilde{\pi}_{An})$ with $p_{in}$ and  $(\ou{\omega}{A}{n},-\ou{\utilde{\omega}}{A}{n})$ with $\ou{q}{i}{n}$. The action of the Hamiltonian vector field of the Raychaudhuri constraint $H[N]$ on these fundamental fields is given in \eref{Lxi-piom} and \eref{Lxi-t-piom}. At the quantum level, the underlying symmetries are generated through the Ward identities \eref{Ward-ident}. The corresponding stress energy tensor is
\begin{align}\nonumber
T(z)=\frac{\pi}{\I\hbar}\Big[\normord{(\partial \pi_A)(z)\omega^A(z)}&-\normord{(\partial \utilde{\pi}_A)(z)\utilde{\omega}^A(z)}+\\
&-\normord{\pi_A(z)(\partial \omega^A)(z)}+\normord{\utilde{\pi}_A(z)(\partial\utilde{\omega}^a)(z)}\Big].\label{T-SU11-def}
\end{align}
Since the $SU(1,1)$-modes $\pi_A$, $\omega^A$ and $\utilde{\pi}_A$, $\utilde{\omega}^A$ each take values in $\C^2$, we obtain $d=2\times 2$, and the central charge is
\begin{equation}
c^{SU(1,1)}=-4.
\end{equation}
On each plaquette, we then have fundamental bosonic Heisenberg relations
\begin{align}
\left[\pi_{An}(i),\omega_{Bm}(j)\right]_+&=-\I\,\hbar\,\epsilon_{AB}\delta_{n+m}\delta_{i,j},\quad i,j=1,\dots,N,\label{pi-om-Hberg1}\\
\left[\utilde{\pi}_{An}(i),\utilde{\omega}_{Bm}(j)\right]_+&=+\I\,\hbar\,\epsilon_{AB}\delta_{n+m}\delta_{i,j},\quad i,j=1,\dots,N\label{pi-om-Hberg2}.
\end{align}
Using $\eta_{A\bar A}\eta_{B\bar B}\bar{\epsilon}^{\bar A\bar B}=-\epsilon_{AB}$, with $\eta_{A\bar{A}}$ denoting the $SU(1,1)$ inner product, see \eref{SU1-1-invrncs}, it easy to check that the reality conditions \eref{realcond1}, \eref{realcond2} are compatible with the Heisenberg commutation relations \eref{pi-om-Hberg1} and \eref{pi-om-Hberg2}. 
\paragraph{Total central charge} We saw that the chronton modes $(p_\chi,\chi)$ can only be bosonic, while the area modes can be either bosons ($+$) or fermions ($-$). For the time being, we assume bosonic commutation relations for the $SU(1,1)$ modes as well. Under this assumption, the total central charge of the auxiliary CFT on each plaquette is
\begin{equation}
c^{\mtext{plaquette}}=c^{(\chi)}+c^{(b)}+c^{SU(1,1)}=2\mp 1-4<0.\label{cntrl-chrg-bsn}
\end{equation}
That the central charge is negative is problematic, for it implies a violation of unitarity \cite{CFTbook}. In the next section, we show how to avoid this problem by assuming fermionic commutation relations for the $SU(1,1)$ modes. This has important physical consequences. We will see that the fermionic representation excludes the ultra-Planckian $L^2< c\bar{c}$ regime.

\subsection{The case for Fermi--Dirac statistics for the $SU(1,1)$ modes}\label{sec3.3}
\noindent Let us now investigate what happens when assuming Fermi--Dirac statistic for the $SU(1,1)$ modes. We constructed these modes from the two eigenvectors $\pi^A$ and $\omega^A$ of  the $SU(1,1)$ momentum $\Pi=JL+c\bar{X}+\bar{c}X$, see \eref{eigen-spinrs}. The off-diagonal elements $c$ and $\bar{c}$ of $\Pi$ describe the shear of the null generators, see \eref{c-def}. The shear is a sort-of non-linear graviton, which we must satisfy bosonic commutation relations. This does not imply, however, that $\pi_A$ and $\omega_A$ must be bosons as well. In fact, a very similar construction appears commonly in CFTs, where one can realize an $\mathfrak{so}(N)$ current algebra in term of $N$ fermions \cite{CFTbook}. 

\paragraph{Evaluation of the Casimir} First of all, we compute the $SU(1,1)$ Casimir in this fermionic representation, taking into account that $\pi_A$ and $\omega_A$ are now Grassmann-valued. Basic spinor algebra, see \cite{penroserindler}, implies
\begin{align}\nonumber
\mathrm{Tr}\left(\Pi^2\right)&=\ou{\Pi}{A}{B}\ou{\Pi}{B}{A}=-\Pi_{AB}\Pi^{AB}=-\pi_{(A}\omega_{B)}\pi^A\omega^B=\\
&=+\frac{1}{2}\pi_A\pi^A\omega_B\omega^B-\frac{1}{2}(\pi_A\omega^A)^2.
\end{align}
In the bosonic case, $\omega_A\omega^A=0$. In the fermionic case, it may be not. Instead, we have
\begin{align}\nonumber
\pi_A\pi^A\omega_B\omega^B&=\epsilon^{AB}\epsilon_{CD}\pi_A\pi_B\omega^C\omega^D=2\delta^{[A}_C\delta^{B]}_D\pi_A\pi_B\omega^C\omega^D
=-2(\pi_A\omega^A)^2\end{align}
On the other hand, we also know
\begin{equation}
\mathrm{Tr}\left(\Pi^2\right)=-2(L^2-c\bar{c}).
\end{equation}
Putting everything together, we find
\begin{equation}
L^2-c\bar{c}=\frac{3}{4}(\pi_A\omega^A)^2.\label{Csmr-w-crrents}
\end{equation}
\paragraph{Reality conditions} 
Next, we consider the reality conditions.  For $\ou{\Pi}{A}{B}$ to lie in $\mathfrak{su}(1,1)$, it is not enough to impose tracelessness, i.e. $\Pi_{[AB]}=\tfrac{1}{2}\epsilon_{AB}\mathrm{Tr}(\Pi)=0$. We need to satisfy the supplementary reality conditions \eref{realcond} as well. 
At the quantum level, complex conjugation turns into hermitian conjugation \qq{$\dagger$}, which reverses the order of operators. In what follows, we denote by $\bar{\pi}_{\bar A}$ and $\bar{\omega}_{\bar A}$ the conjugate of $\pi_{A}$ and $\omega_{A}$ respectively, with barred indices referring to complex conjugation,
\begin{equation}
\bar{\pi}_{\bar A}:=[\pi_A]^\dagger,\qquad\bar{\omega}_{\bar A}:=[\omega_A]^\dagger.
\end{equation}
This simplifies our notation moving forward. Taking into account $\xi_A\psi^A=-\xi^A\psi_A$, which is a consequence of the anti-symmetry of the epsilon tensor $\epsilon_{AB}=-\epsilon_{BA}$, we obtain
\begin{equation}
\pi_{(A}\omega_{B)}\stackrel{!}{=}-\eta_{A\bar A}\eta_{B\bar B}\bar{\omega}^{(\bar B}\bar{\pi}^{\bar A)}=+\eta_{A\bar A}\eta_{B\bar B}{\bar{\pi}}^{(\bar B}{\bar{\omega}}^{\bar A)},
\end{equation}
where $\eta_{A\bar{A}}$ is the signature $(-$$+)$ inner product, which is invariant under $SU(1,1)$.\footnote{We can take, for concreteness, a representation where $\eta_{0\bar{0}}=-1$, $\eta_{1\bar{1}}=+1$, $\epsilon_{01}=\epsilon^{01}=-\epsilon^{10}=-\epsilon_{10}=1$, and all other entries of $\eta_{A\bar A}$ and $\epsilon_{AB}$ vanish.}\smallskip

The eigenspinors $\pi_A$ and $\omega_A$ are unique up to exchange $(\pi,\omega)\rightarrow(-\omega,\pi)$ and global rescalings $(\pi_A,\omega_A)\rightarrow(\E^z\pi_A,\E^{-z}\omega_A)$ with $z\in\C$. Thus, there are now two cases to distinguish.\smallskip

\begin{enumerate}
\item In the first case, we match $\pi_A$ with $\bar{\pi}_{\bar A}$ and $\omega_A$ with $\bar{\omega}_{\bar A}$. Thus, there exists some unspecified $z\in\C$ such that
\begin{equation}
\pi_A=\E^{+z}\eta_{A\bar A}\bar{\pi}^{\bar A},\qquad \omega_A=\E^{-z}\eta_{A\bar A}\bar{\omega}^{\bar A}.\label{spinr-match}
\end{equation}
A short calculation gives
\begin{align}
\pi_A&=\E^{+z}\eta_{A\bar A}\bar{\pi}^{\bar{A}}=-\E^{+z+\bar{z}}\eta_{A\bar{B}}\eta^{B\bar{B}}\pi_B=\E^{+z+\bar{z}}\pi_A,\\
\omega_A&=\E^{-z}\eta_{A\bar A}\bar{\omega}^{\bar{A}}=-\E^{-z-\bar{z}}\eta_{A\bar{B}}\eta^{B\bar{B}}\omega_B=\E^{-z-\bar{z}}\omega_A.
\end{align}
This implies $z=-\bar{z}$. We can now rescale $\pi_A$ and $\omega_A$ by a mere phase. In this way, it is always possible to attain
\begin{equation}
\pi_A=\eta_{A\bar A}\bar{\pi}^{\bar A},\qquad \omega_A=\eta_{A\bar A}\bar{\omega}^{\bar A}.\label{f-realcond1}
\end{equation}
This implies
\begin{equation}
\bar{\omega}^{\bar A}\bar{\pi}_{\bar{A}}=-\bar{\omega}^{\bar{B}}\eta_{B\bar{B}}\eta^{B\bar{A}}\bar{\pi}_{\bar A}=\bar{\omega}^{\bar{B}}\eta_{B\bar{B}}\ou{\eta}{B}{\bar A}\bar{\pi}^{\bar A}=\omega_A\pi^A=\pi_A\omega^A.
\end{equation}
In our notation, this is the same as to say,
\begin{equation}
(\pi_A\omega^A)^\dagger =\pi_A\omega^A.\label{piom-realty}
\end{equation}
Going back to the expression for the $SU(1,1)$ Casimir in terms of the fermionic currents, i.e.\ \eref{Csmr-w-crrents}, we obtain
\begin{equation}
L^2\geq c\bar{c}.
\end{equation}
Notice also that for each mode, the reality conditions \eref{f-realcond1} translate into
\begin{equation}
\pi_{An}(i)=\eta_{A\bar{A}}\ou{\bar{\pi}}{\bar{A}}{-n}(i),\qquad\omega_{An}(i)=\eta_{A\bar{A}}\ou{\bar{\omega}}{\bar{A}}{-n}(i),\label{f-realcond-plaqtt1}
\end{equation}

\item In the second case, we match $\pi_A$ with $\bar{\omega}_{\bar A}$ and $\omega_A$ with $\bar{\pi}_{\bar A}$. However, we now need to reverse the relative order of the eigenspinors. Compared to \eref{spinr-match}, we thus need to take into account an additional minus sign when matching the fields,
\begin{equation}
\pi_A=\E^{+z}\eta_{A\bar A}\bar{\omega}^{\bar A},\qquad \omega_A=-\E^{-z}\eta_{A\bar A}\bar{\pi}^{\bar A},
\end{equation}
where $z$ is some yet unspecified complex number. A short calculation gives
\begin{align}
\pi_A&=\E^{+z}\eta_{A\bar A}\bar{\omega}^{\bar{A}}=+\E^{+z-\bar{z}}\eta_{A\bar{B}}\eta^{B\bar{B}}\omega_B=-\E^{+z-\bar{z}}\pi_A,\\
\omega_A&=-\E^{-z}\eta_{A\bar A}\bar{\pi}^{\bar{A}}=+\E^{\bar{z}-z}\eta_{A\bar{B}}\eta^{B\bar{B}}\omega_B=-\E^{\bar{z}-z}\omega_A.
\end{align}
This implies $z-\bar{z}=\pm\I\pi$. Thus $z=2r\pm \I\frac{\pi}{2}$. We can now rescale $\pi_A$ and $\omega$ by the dilatation $\pi_A\rightarrow \E^{-r}\pi_A$ and $\omega_A\rightarrow \E^r\omega_A$. In this way, it is always possible to achieve
\begin{equation}
\pi_A=\pm\I\eta_{A\bar A}\bar{\omega}^{\bar A},\qquad \omega_A=\mp\I\eta_{A\bar A}\bar{\pi}^{\bar A}.\label{f-realcond2}
\end{equation}
This implies
\begin{equation}
\bar{\omega}^{\bar A}\bar{\pi}_{\bar{A}}=-\bar{\omega}^{\bar{A}}\eta_{B\bar{A}}\eta^{B\bar{B}}\bar{\pi}_{\bar B}=\bar{\omega}^{\bar{A}}\eta_{B\bar{A}}\ou{\eta}{B}{\bar{B}}\bar{\pi}^{\bar{B}}=\omega_B\pi^A=\pi_A\omega^A.
\end{equation}
This is the same condition we encountered before, see \eref{piom-realty}. Thus, in both cases, we have
\begin{equation}
L^2\geq c\bar{c}.
\end{equation}
Notice also that for each mode, the reality conditions \eref{f-realcond1} translate into
\begin{equation}
\pi_{An}(i)=\pm\I\eta_{A\bar{A}}\ou{\bar{\omega}}{\bar{A}}{-n}(i),\qquad\omega_{An}(i)=\mp\I\eta_{A\bar{A}}\ou{\bar{\pi}}{\bar{A}}{-n}(i),\label{f-realcond-plaqtt2}
\end{equation}
 \end{enumerate}
In either case, we will have fermionic anti-commutation relations
\begin{align}
\left[\pi_{An}(i),\omega_{Bm}(j)\right]_-&=-\I\,\hbar\,\epsilon_{AB}\delta_{n+m}\delta_{i,j},\quad i,j=1,\dots,N,\label{f-pi-om-Hberg1}\\
\left[\utilde{\pi}_{An}(i),\utilde{\omega}_{Bm}(j)\right]_-&=+\I\,\hbar\,\epsilon_{AB}\delta_{n+m}\delta_{i,j},\quad i,j=1,\dots,N\label{f-pi-om-Hberg2}.
\end{align}
 Using $\eta_{A\bar A}\eta_{B\bar B}\bar{\epsilon}^{\bar A\bar B}=-\epsilon_{AB}$, with $\eta_{A\bar{A}}$ denoting the $SU(1,1)$ inner product, see \eref{SU1-1-invrncs}, it easy to check that the reality conditions \eref{f-realcond-plaqtt1}, \eref{f-realcond-plaqtt1} are compatible with the Heisenberg commutation relations \eref{f-pi-om-Hberg1} and \eref{f-pi-om-Hberg2}. 
 \paragraph{Total central charge of the fermionic model} We saw previously that the chronton modes $(p_\chi,\chi)$ can only be bosonic, but for the area modes $b$ and $\bar{b}$ and the $SU(1,1)$ modes we can choose either bosonic or fermionic commutation relations. If we assume that both the area modes as well as the $SU(1,1)$ modes are fermionic (a mixed case seems somewhat odd), the total central charge of the model turns out to be
 \begin{equation}
c^{\mtext{plaquette}}=c^{(\chi)}+c^{(b)}+c^{SU(1,1)}=2+ 1+4=7.\label{cntrl-chrg-frmi}
\end{equation}
If the central charge is negative, the CFT is non-unitary. If the central charge is positive, we are on much safer grounds to define a positive-definite inner product. On positive norm states, we would then always have
\begin{equation}
L^2\geq c\bar{c}.\label{lum-bnd}
\end{equation}
To conclude, we discuss the physical significance of this observation below.

\section{Unitarity implies Planck luminosity bound}\label{sec4}
\noindent
Imagine an idealized asymptotic observer  in an  asymptotically flat spacetime.\footnote{Operationally, an asymptotic boundary can only be an approximation to a more realistic notion of \emph{finite infinity}, see \cite{Ellis1984,Wiltshire:2007jk} and also \cite{Ashtekar:2024xwl} for a similar viewpoint. } Provided the matter fields satisfy the dominant energy condition, the ADM (Arnowitt--Deser--Misner) mass \cite{ADMmass} is always positive \cite{Schon:1979rg,Wittenproof}. Furthermore, not only is the mass positive, also its flux will have a definite sign \cite{Bondi21,Sachs103}. 
The flux can be quantified in terms of the Bondi mass loss formula
\begin{equation}
\dot{M}_{\mathrm{B}}(u)=-\frac{1}{4\pi G}\oint_{S^2_u\subset{\mathcal{I}_+}}d^2v_o\,\dot{\sigma}^{(0)}\dot{\bar{\sigma}}^{(0)}.\label{Bondi-lss}
\end{equation}
In here, $\dot{{\sigma}}^{(0)}(u,\zeta,\bar{\zeta})$ is the time derivative of the asymptotic Bondi shear and the integral is evaluated at a two-dimensional cut of future null infinity at constant Bondi time $u$. In addition, $\zeta$ labels points on the sphere and $d^2v_o=\sin\vartheta\,\di\theta\wedge\di\varphi$ is the standard (round) surface area element.\smallskip

In a region far away from all sources, we can then integrate Einstein's equations. Using a double null foliation with co-normals $k_a=-\nabla_a u$, $\ell_a\propto\nabla_a\rho$, in which $k^a=\partial^a_r$ is an outgoing affine null direction, which lies transversal ($k_a\ell^a=-1$) to the  $\rho=\mathrm{const}.$ null hypersurfaces $\mathcal{N}_\rho$, which are approaching future null infinity as $\rho\rightarrow\infty$ ($\lim_{\rho\rightarrow\infty}\mathcal{N}_\rho\subset\mathcal{I}^+$, $\rho=\mathcal{O}(r)$), we obtain, see e.g. \cite{Wieland:2020gno},    
\begin{align}
\sigma_{(\ell)}(u,r,\zeta,\bar{\zeta})&=-\frac{\dot{\sigma}^{(0)}(u,\zeta,\bar{\zeta})}{r}+\mathcal{O}(r^{-2}),\\
\vartheta_{(\ell)}(u,r,\zeta,\bar{\zeta})&=-\frac{2}{r}+\mathcal{O}(r^{-1}).
\end{align}
The Bondi formula determines the total radiated power. In physical units (watts), the power radiated per fixed solid angle is
\begin{equation}
\mathcal{L}_{\mathrm{B}}(u,\zeta,\bar{\zeta})=\frac{4c^5}{G}\lim_{r\rightarrow\infty}\frac{\bar{\sigma}_{(\ell)}(u,r,\zeta,\bar{\zeta})\sigma_{(\ell)}(u,r,\zeta,\bar{\zeta})}{(\vartheta_{(\ell)}(u,r,\zeta,\bar{\zeta}))^2}.
\end{equation}
The pre-factor is the Planck power. This power is reached in processes that radiate one Planck unit of energy in one unit of Planck time, 
\begin{equation}
\frac{m_{\mathrm{P}}c^2}{t_{\mathrm{P}}}=\frac{c^5}{G}\approx \SI{E52}{\watt}.
\end{equation}
In our model, we have a quantisation of shear and expansion at finite distance. Instead of infinitely many modes on the sphere, we have a tesselation into $N$ plaquettes $\{\mathcal{D}_i\}_{i=1}^N$. On every such plaquette, there is a $U(1)$ generator $L$, see \eref{L-def}. This $U(1)$ generator determines the expansion of a light ray that emanates from the plaquette. The relation between the algebraic $SU(1,1)$ generator $L$ and the geometric observables, such as expansion and area, is
\begin{equation}
L^2=\frac{1}{(16\pi\gamma G)^2}(\Omega^2\vartheta)^2,
\end{equation}
where $\Omega^2$ is the conformal factor. Furthermore, we also have a representation of the shear of such a null ray in terms of the off-diagonal components $c$ and $\bar{c}$ of the $\mathfrak{su}(1,1)$-valued momentum density, see \eref{Pi-def} and \eref{c-def}. We obtain
\begin{equation}
c\bar{c}=\frac{1}{(8\pi\gamma G)^2}(1+\gamma^2)\Omega^4\sigma\bar{\sigma}.
\end{equation}
In the last section, we saw that each light ray carries a CFT. This CFT has a central charge, which is a sum of three terms \eref{cntrl-chrg-frmi}. To exclude otherwise unavoidable negative norm states, the central charge must be positive. Otherwise, we have a non-unitarity CFT. A unitary CFT is only possible if
\begin{equation}
L^2\geq c\bar{c}.
\end{equation}
This implies
\begin{equation}
\vartheta^2\geq 4(1+\gamma^2)\sigma\bar{\sigma},
\end{equation}
which translates into the luminosity bound
\begin{equation}
\mathcal{L}_{\mathrm{B}}\leq\mathcal{L}_{\mtext{crit}.}=\frac{c^5}{G}\frac{1}{1+\gamma^2}.
\end{equation}

\section{Summary and discussion}\label{sec5}

\paragraph{Summary} Here we proposed a new non-perturbative quantisation of gravitational null initial data. The proposal is developed in three steps. \emph{First}, we considered the {classical} phase space of  local gravitational subsystems whose bulk and boundary dynamics is governed by the $\gamma$-Palatini--Holst action \cite{holst,Parviol,surholst}. We studied such subsystems in light cone gauge. The subsystems in question are characterized by the gravitational null initial data on a compact region $\mathcal{N}\cong[0,2\pi]\times S^2$  with two disconnected spacelike boundaries $\partial \mathcal{N}=\mathcal{C}_+\cup\mathcal{C}_-$ on an abstract three-dimensional null surface. The resulting physical phase space is composed of the two radiative modes that can cross the null surface and additional gravitational boundary (edge) modes \cite{Balachandran:1995qa,PhysRevD.51.632,Donnelly:2016auv,Donnelly:2016rvo,Gomes:2016mwl,Geiller:2017xad,Speranza:2017gxd,Geiller:2017whh,Wieland:2017cmf,Wieland:2017zkf,Takayanagi:2019tvn,Freidel:2019ees,Francois:2021aa,Freidel:2020ayo,Freidel:2020svx,Freidel:2020xyx,Donnelly:2020xgu,Freidel:2021cjp,Carrozza:2021gju,Carrozza:2022xut,Kabel23,Giesel:2024xtb,Ciambelli:2022vot,Wieland:2021vef,Araujo-Regado:2024dpr} at $\mathcal{C}_\pm$. We built this classical phase space using standard symplectic methods. The starting point is an auxiliary kinematical phase space. There are three types of kinematical fields: the \emph{chronoton modes} $(p_\chi,\chi)$ that determine a physical time coordinate $\mathcal{U}$ relative to an auxiliary and unphysical clock $u$, complex-valued \emph{area modes} $(b,\bar{b})$ that satisfy standard oscillator Poisson brackets, and additional $SU(1,1)$ modes that encode the \emph{shape degrees of freedom} of the two-dimensional geometry of the spatial cuts of $\mathcal{N}$. At the quantum level, the chronoton provides, in fact, nothing but a quantum reference frame \cite{Rovelli:1990pi,Giacomini:2017zju,Loveridge2018,Vanrietvelde:2018pgb,Hoehn:2019owq,delaHamette:2021oex}
for angle-dependent reparametrizations of the null coordinate. In addition, there is an area edge mode $(a,\bar{a})$ that satisfies standard oscillator Poisson brackets. At the quantum level, the corresponding number operator determines the area of an arbitary cross section of $\mathcal{N}$.\footnote{The resulting area spectrum agrees with earlier results from loop quantum gravity \cite{Rovelliarea,AshtekarLewandowskiArea,Haggard:2023tnj,bianchisommer} for the area flux \cite{FernandoBarbero:2009ai,Beetle:2010rd,Krasnov:1998mp,Wieland:2017cmf} operator in the spin network representation \cite{status, thiemann, rovelli,Ashtekar:2021kfp}.}  Here, we took this to be the final cut $\mathcal{C}_+$ of $\mathcal{N}$.  The construction of the kinematical phase space was concluded by introducing a free-fermion (bosonic) representation of the $SU(1,1)$ shape modes.  Within the kinematical phase space, there lies the physical phase space, which is the quotient of the constraint hypersurface by the group of local gauge symmetries. The gauge symmetries are generated by three first-class constraints. Besides the first-class constraints, there are four second-class constraints. We have not given the Dirac bracket explicitly, but the Dirac matrix has a simple block skew-symmetric structure \eref{Up-Phi-brck}, \eref{Phi-Xi-brck}, \eref{Xi-bXi-brck} and can be immediately inverted using the skew-symmetric version of the Heaviside step function, which is the inverse of $\dot{\delta}(u-u')$. The kinematical fields are charged under a boundary symmetry algebra generated by the first-class constraints. The fundamental constraints are the Hamiltonian constraint, which generates angle-dependent reparametrizations of the null generators, and two additional scalar constraints generating abelian $U(1)$ transformations of the kinematical boundary variables. For ultra-Planckian modes, one of these local $U(1)$ transformations is replaced by the real multiplicative group $\R_+$. The Hamiltonian constraint is nothing but the Raychaudhuri equation, which is the $R_{ab}\ell^a\ell^b=0$ component of Eisntein's equations.\smallskip

\emph{Second}, we took the analysis to the quantum level. Following earlier developments on isolated horizons and generic null boundaries \cite{Ashtekar:2000eq,isohorizon,Ashtekar:1999wa,Ashtekar:aa,BarberoG.:2012ae,DiazPoloPranzetti,Ashtekar:2001is,Engle:2010kt} and the quantisation of gravitational edge modes \cite{Wieland:2017cmf,Wieland:2017ksn,Freidel:2020ayo}, we considered a polymer quantisation \cite{shadowstats} in which the two-dimensional cuts of the null surface are tessellated into finitely many plaquettes.\footnote{This implies that so-called superrotations will no longer be weakly continuous, thus replacing $\mathrm{Diff}(S_2)$ by the permutation group of $N$ punctures.}
In this way, the infinitely many $Y_{\ell m}$ angular modes are replaced by finitely many punctures on the cut. By using appropriate smearing functions and integrating the kinematical variables, each of which carries a definite density weight, over each plaquette, we introduced a regularization of the classical Heisenberg algebra. At the same time, we kept the infinitely many modes along each null ray intact. Using a standard Fourier series expansion into positive and negative frequency modes, we introduced the resulting Fock vacuum. To obtain a realization of the Hamiltonian constraint on this Hilbert space, we considered an auxiliary CFT akin to the $\beta\gamma$ ($bc$) system of string theory. To start out, we left the underlying statistics unspecified. The CFT was selected by matching its field content and commutation relations with the mode expansion and Fock space representation that we introduced earlier. The perhaps single most crucial element that distinguishes our results from earlier developments in the area \cite{Fuchs:2017jyk,Ciambelli:2024swv,Ciambelli:2023mir}  is the inclusion of the Barbero--Immirzi parameter, which allows us to compactify otherwise unbounded directions (e.g.\ dilations of the null normal) of the classical phase space.\footnote{In the earlier literature, this was understood as a mere change of canonical variables \cite{Barbero1994,Immirziparam}.} This in turn implies that new representations can appear in the spectrum of the quantum theory that otherwise do not. For example, in our case, the cross sectional area and the expansion of the null generators arise from CFT currents, which are quadratic in terms of the fundamental oscillator type variables. These oscillators are charged under a local $U(1)$ gauge symmetry. The discrete spectra of geometric operators that we see in loop quantum gravity are a result of the compactness of these gauge orbits. None of this happens when $\gamma\rightarrow\infty$, which is the limit in which the Barbero--Immirzi $\gamma$ parameter disappears from the classical action \eref{acnt-def}. In this limit, the $U(1)$ generator \eref{L-def} always vanishes, thus restricting ourselves to representations in which $L^2-c\bar{c}\leq 0$. 

Upon having identified an auxiliary conformal field theory whose currents (the stress energy tensor $T(z)$ and the $U(1)$ current $j(z)$) can be matched to their gravitational analogues on the null cone, we computed the central charge. We considered both free fermionic and bosonic representations of the $SU(1,1)$ current algebra. For bosonic representations, $L^2-c\bar{c}$ can have either sign. For fermionic representations, the central charge is always positive and $L^2-c\bar{c}$ is always greater or equal to zero.
\smallskip

\emph{Third and finally}, we returned to an earlier conjecture on gravitational wave luminosity that we made in \cite{Wieland:2025LP}. The basic idea is to derive a quantum bound on the radiated power for gravitational waves from physical requirements on the central charge of the auxiliary CFT. Theories in which the central charge is negative violate unitarity. Upon excluding all states in which the central charge is negative,  we were able to conclude that the $SU(1,1)$ Casimir $L^2-c\bar{c}$ is always greater or equal to zero. To connect this quasi-local bound to asymptotic observables, we considered a double null foliation in the vicinity of future null infinity, see e.g.\  \cite{Wieland:2020gno}.  We wrote the asymptotic Bondi flux as the square modulos of the ratio of the asymptotic shear and the asymptotic expansion. Upon taking the $r\rightarrow \infty$ limit, we recovered the bound reported earlier \cite{Wieland:2025LP}.

\paragraph{Open problems} There are many open problems. Let us single out two of them. \hyperref[sec4]{Section 4} was largely based on semi-classical assumptions. What is missing is to justify these assumptions by an appropriate choice of boundary states. This could be probably best achieved by introducing an auxiliary $\rho$-parameter family of coherent states $\{\Psi_\rho\}_{\rho\in\R_+}$. This family of states would be  then chosen such that they can characterize semi-classical null initial data on a family of null hypersurfaces $\mathcal{N}_\rho$ that lie within the \emph{same} classical solution $(\mathcal{M}, g_{ab})$ to Einstein's equations.
In addition, the hypersurfaces $\mathcal{N}_\rho$ should be chosen such that they foliate a spacetime region $\coprod_\rho\mathcal{N}_\rho\subset\mathcal{M}$ extending all the way to the asymptotic future (past) null boundary $\mathcal{I}_\pm$. We have not constructed such coherent states explicitly. However, we have a simple oscillator representation at hand, and we expect that the corresponding vertex operators are good candidates for such geometrical coherent states. At the level of the discrete spin network representation of quantum geometry, there is a large class of such  coherent states avalilable \cite{Freidel:2010tt,Freidel:2010aq,Sahlmann:2001nv,Thiemann:2000bw}. Another important  problem that we only briefly mentioned is the issue of caustics. In our entire construction, we assumed the absence of light rays that could leave the hypersurface. At this point, we can only speculate. We believe there are  three clear paths for future research ahead: First, following a second quantised approach akin to group field theory (GFT) in which the wave functional for a single light ray is promoted into a field operator. In such a scenario light rays could be created through non-trivial interaction vertices in the auxiliary GFT action. The second (and probably related) option is to extend the auxiliary CFT path integral allowing for a sum over two-dimensional topologies, thus mirroring the sum over worldsheets in string theory. The third and most conservative approach is based on coarse graining techniques developed for the spinfoam state sum models of quantum gravity in which we have appropriate refinement maps to take the $N\rightarrow\infty$ projective limit. The possible presence of caustics in the resulting quantum light-cone geometry could be then visible in the entanglement structure of the underlying quantum geometry \cite{Bianchi:2012ev,Cao:2016mst} encoding adjacency (and breaking of adjacency) between neighbouring light rays. 

\paragraph{Acknowledgments} I would like to thank Abhay Ashtekar, Eugenio Bianchi, Laurent Freidel and Simone Speziale for discussions. I thank the organizers of the QISS 2025 conference in Vienna for the opportunity to present this research. This research was funded through a Heisenberg fellowship of \emph{Deutsche Forschungsgemeinschaft} (DFG, German Research
Foundation)—543301681. In addition, this research was supported in part by \emph{Österreichischer Wissenschaftsfonds FWF}  (Austrian Science Fund) through the emerging fields programme, \doi{10.55776/EFP6}.
\section*{Appendix}\label{appdx}
\setcounter{section}{1}
\setcounter{equation}{0}
\renewcommand{\theequation}{\Alph{section}.\arabic{equation}}
\noindent In here, we provide a brief derivation of the gravitational pre-symplectic potential for the $\gamma$-Holst action on a null surface. For simplicity, we ignore any boundary terms in the action. A derivation in terms of a  boundary Lagrangian can be found in \cite{Wieland:2021vef}. 

To start out, we consider the selfdual decomposition \cite{selfdualtwo,komplex1,Wieland:2013cr} of the $\gamma$-Palatini action \cite{holst, Parviol} in a compact spacetime region $\mathcal{M}$
\begin{equation}
S_{\mathcal{M}}[e,\omega]=\left[\frac{\I}{8\pi \gamma G}(\gamma+\I)\int_{\mathcal{M}}\Sigma_{AB}\wedge F^{AB}\right]+\CC\label{acnt-def}
\end{equation}
In here, $\Sigma_{AB}=\Sigma_{BA}$ is the Pleba\'nski selfdual two-form, i.e.\ the selfdual part of the bivector $\Sigma_{\alpha\beta}=e_\alpha\wedge e_\beta$, and $F_{AB}=F_{BA}$ is the selfdual curvature,
\begin{align}
\Sigma_{AB}&=-\frac{1}{2}e_{A\bar{C}}\wedge \tensor{e}{_B^{\bar{C}}},\\
\ou{F}{A}{B}&=\di \ou{\omega}{A}{B}+\ou{\omega}{A}{C}\wedge\ou{\omega}{C}{B}.
\end{align}
Given the action, the resulting field equations are $\nabla\Sigma_{AB}=0$ and $F_{AB}\wedge\tensor{e}{^B_{\bar{A}}}=0$. The first set of equations constrains the connection: if the tetrad $e_{A\bar{A}}$ is non-degenerate,\footnote{This is the same as to say that the four-volume is non-vanishing, i.e.\ $\Sigma_{AB}\wedge\Sigma^{AB}\neq 0$.} it is equivalent to the torsionless condition $\nabla e_{A\bar{A}}=0$. The second set of equations, i.e.\ $F_{AB}\wedge\tensor{e}{^B_{\bar{A}}}=0$, are equivalent to the usual Einstein's equations (plus Bianchi identity $\nabla^2e_{A\bar A}=0$). Notice that the Barbero--Immirzi parameter $\gamma$ does not affect the field equations. As we will see below, and it has been noted many times in the literature \cite{Wieland:2017cmf,Wieland:2021vef,Freidel:2020ayo,Freidel:2020xyx,Giesel:2024xtb} before, the Barbero--Immirzi parameter $\gamma$ does change, however, the symplectic potential and definition of charges. 

The variation of the action contains a boundary term. This term determines the pre-symplectic current. The integral of this current along a partial Cauchy surface $\mathcal{N}$, which we take to be null, defines the pre-symplectic potential
\begin{equation} 
\Theta^{\mtext{Holst}}_{\mathcal{N}}=\left[\frac{\I}{8\pi \gamma G}(\gamma+\I)\int_{\mathcal{N}}\Sigma_{AB}\wedge \bbvar{d}\omega^{AB}\right]+\CC,\label{app:theta-N1}
\end{equation}
for the quasi-local phase space on $\mathcal{N}$.
The integrand depends only on the pull-back of the differential forms $\Sigma_{AB}$ and $\bbvar\omega_{AB}$ to $\mathcal{N}$. As explained in \cite{Wieland:2019hkz,Wieland:2017zkf}, the pull-back of $\Sigma_{AB}$ to a null hypersurface satisfies
\begin{equation}
\varphi^\ast_{\mathcal{N}}\Sigma_{AB}=\ell_{(A}e_{B)}\wedge\bar{m}.\label{app:e-l-def}
\end{equation}
In here, $e_A$ is a boundary intrinsic spinor-valued one-form
\begin{equation}
e_A=(\ell_A k-k_A{m})\,\operatorname{mod}\,\bar{m}.
\end{equation}
In addition, $(k_A,\ell_A)$ is a spin dyad. Given a spin dyad, we also have a Newman--Penrose tetrad $(\I\ell_A\bar{\ell}_{\bar{A}},\I k_A\bar{k}_{\bar{A}},\I\ell_A\bar{k}_{\bar{A}},\I k_A\bar{\ell}_{\bar{A}})\equiv({}^4k_a,{}^4\ell_a,{}^4m_a,{}^4\bar{m}_a)$ adapted to $\mathcal{N}$. It is adapted to $\mathcal{N}$, because (in our conventions) the pull-back of ${}^4\ell_a$ (rather than ${}^4k_a$) to the null surface vanishes. The pull-back of ${}^4k_a$ to $\mathcal{N}$, on the other hand, defines the so-called Ehresmann connection \cite{Freidel:2024emv} $k_a$ on $\mathcal{N}$ and $m_a=\varphi^\ast_{\mathcal{N}}{}^4m_a$ is the co-dyad \eref{q-dyad}. It may seem a little bizarre at this point to use similiar symbols, e.g.\ $k_a$ and $k_A$, for both a co-vector and a spinor, but context will always clarify notation. In here, $k$ or $k_a$ always denotes a one-form on $\mathcal{N}$, capital indices $A,B,C,\dots$ label spinors. \smallskip

The spin dyad $(k_A,\ell_A)$ is normalized to $k_A\ell^A=\epsilon_{BA}k^B\ell^A=1$. Any such spin dyad can always be parametrized in terms of some reference dyad $(\mathring{k}_A,\mathring{\ell}_A)$, fixed by some auxiliary gauge condition, and a local $SL(2,\C)$ gauge  transformation $\ou{\Lambda}{A}{B}$. We thus have
\begin{equation}
k^A=\ou{\Lambda}{A}{B}\mathring{k}^B,\qquad \ell^A=\ou{\Lambda}{A}{B}\mathring{\ell}^B.
\end{equation}
Since $(\mathring{k}_A,\mathring{\ell}_A)$ are kept fixed, the variations satisfy
\begin{equation}
\bbvar{d}k^A=-(\bbvar{d}\ou{\Lambda}{A}{C})\uo{\Lambda}{B}{C}k^C,\qquad
\bbvar{d}\ell^A=-(\bbvar{d}\ou{\Lambda}{A}{C})\uo{\Lambda}{B}{C}\ell^C.
\end{equation}
This equation suggests to define the one-form on field space $\ou{{\bbgreek{\Lambda}}}{A}{B}:=$$-(\bbvar{d}\ou{\Lambda}{A}{C})\uo{\Lambda}{B}{C}=\ou{[(\bbvar{d}\Lambda)\Lambda^{-1}]}{A}{B}$, which defines a field space connection with vanishing curvature \cite{Gomes:2019xto,Gomes:2016mwl}. The corresponding covariant field space exterior derivative is $\bbvar{D}$. It acts on the fundamental fields as
\begin{align}
\bbvar{D}e_{A\bar{A}}&:=\bbvar{d}e_{A\bar{A}}-\ou{{\bbgreek{\Lambda}}}{B}{A}e_{B\bar{A}}-\ou{\bar{\bbgreek{\Lambda}}}{\bar{B}}{\bar{A}}e_{A\bar{B}},\\
\bbvar{D}\ou{\omega}{A}{B}&:=\bbvar{d}\ou{\omega}{A}{B}-\nabla\ou{\bbgreek{\Lambda}}{A}{B}.
\end{align}
We insert this equation back into \eref{app:theta-N1}. Taking into account also  the algebraic form of the Pleba\'{n}ski two-form on the null surface, i.e.\ equation \eref{app:e-l-def}, we obtain
\begin{align}
\Theta^{\mtext{Holst}}_{\mathcal{N}}&=-\left[\frac{\I}{8\pi\gamma G}(\gamma+\I)\oint_{\mathcal{C}_+\cup\mathcal{C}_-^{-1}}e_A\wedge\bar{m}\bbvar{d}\ell^A+\right.\nonumber\\
&\quad+\left.\frac{\I}{8\pi\gamma G}(\gamma+\I)\int_{\mathcal{N}}\left(k\wedge\bar{m}\wedge\bbvar{D}(\ell_AD\ell^A)
-m\wedge\bar{m}\wedge\bbvar{D}(k_AD\ell^A)\right)\right]+\CC,\label{app:theta-N2}
\end{align}
where $D=\varphi^\ast_{\mathcal{N}}\nabla$ and $\partial\mathcal{N}=\mathcal{C}_+\cup\mathcal{C}_-^{-1}$.  The first line defines a boundary symplectic structure 
\begin{equation}
\theta_{\mathcal{C}_\pm}=\int_{\mathcal{C}_\pm}\pi_A\bbvar{d}\ell^A+\CC
\end{equation}
for the spinor-valued edge mode $\ell^A$ and its conjugate momentum, which is given by
\begin{equation}
\pi_A=\frac{\I}{8\pi\gamma G}(\gamma+\I)\varphi_{\mathcal{C}_{\pm}}^\ast(\bar{m}\wedge e_A).
\end{equation}
The quantisation of the resulting boundary symplectic structure was carried out in \cite{Wieland:2017cmf}. It reproduces the \emph{discrete area spectrum} of the cross-sectional area-flux of  loop quantum gravity \cite{FernandoBarbero:2009ai,Beetle:2010rd,Krasnov:1998mp,Wieland:2017cmf}. The second line of \eref{app:theta-N2} defines the symplectic structure for the radiative modes. 
Throughout this paper, we promote the null generators of $\mathcal{N}$ into a universal background structure shared between different spacetime geometries. This amounts to a gauge fixing in which all variations on field space of the null generators satisfy $\bbvar{d}\ell^a\propto \ell^a$. This simplifies the third term of \eref{app:theta-N2}, which becomes
\begin{equation}
\int_{\mathcal{N}}m\wedge\bar{m}\wedge\bbvar{D}(k_AD\ell^A)=-\int_{\mathcal{N}}m\wedge\bar{m}\wedge\bbvar{D}(k\,k_A\ell^aD_a\ell^A).
\end{equation}
Thus, only the $\ell^a$-components of $k_AD_a\ell^A$ enter the symplectic potential. Taking into account the torsionless condition on $\mathcal{N}$, we now also have
\begin{align}
k_A\ell^aD_a\ell^A & = \frac{1}{2}(\kappa_{(\ell)}+\I\varphi_{(\ell)}),\label{app:kDl}\\
\ell_AD\ell^A &= -\frac{1}{2}\vartheta_{(\ell)}m-\sigma_{(\ell)}\bar{m},\label{app:lDl}
\end{align}
 where the spin coefficients $(\kappa_{(\ell)},\varphi_{(\ell)},\vartheta_{(\ell)},\sigma_{\ell})$ denote, in that order, the affinity of the null generators, the $U(1)$ connection coefficient, the expansion and shear, see \eref{Lie-m}, \eref{lDl-kappa-def}.
 
Except for the implicit condition that it be future directed, we have left the null vector $\ell^a\in T\mathcal{N}$ unspecified. To parametrize the space of such null directions, we pick a unique representative $\mathring{\ell}^a$. In here, we identify it with the vector field $\partial^a_{\mathcal{U}}$ that we introduced in \eref{l-U-def} above. Thus,
\begin{equation}
\mathring{\ell}^a=\partial^a_{\mathcal{U}},
\end{equation}
We can then always parametrize the generic null direction $\ell^a$ in terms of $\mathring{\ell}^a$ and a dilation. We define
\begin{equation}
 \ell^a = \E^\lambda\mathring{\ell}^a,\qquad k_a =-\E^{-\lambda}\mathring{k}_a\label{app:null-dilat}.
 \end{equation}
This implies
\begin{equation}
\mathring{k}\wedge m\wedge\bar{m}=-\di\mathcal{U} \wedge m\wedge\bar{m}.
\end{equation}
The spin coefficients transform under such dilations, as
\begin{equation}
({\kappa}_\smallcirc,{\varphi}_\smallcirc,{\vartheta}_\smallcirc,{\sigma}_\smallcirc):=(\kappa_{(\mathring{\ell})},\varphi_{(\mathring{\ell})},\vartheta_{(\mathring{\ell})},\sigma_{(\mathring{\ell})})=\E^{-\lambda}
\left(\kappa_{({\ell})}-\mathcal{L}_\ell\lambda,\varphi_{({\ell})},\vartheta_{({\ell})},\sigma_{({\ell})}\right),\label{app:spincoeff}
\end{equation}
see \eref{kappa-def} and \eref{Lie-m}.
It is easy to check that the contribution to the symplectic potential from the null dilatation \eref{app:null-dilat} is a mere boundary term $\propto\int_{\partial\mathcal{N}}\varepsilon\,\bbvar{d}\lambda$. 

Next, we insert this parametrisation back into the symplectic potential \eref{app:theta-N2}. Taking into account \eref{app:kDl} and \eref{app:lDl}, we obtain
\begin{align}\nonumber
\Theta^{\mtext{Holst}}_{\mathcal{N}}&=\theta_{\mathcal{C}_+}-\theta_{\mathcal{C}_-}-\frac{1}{16\pi G}\int_{\partial\mathcal{N}}\bbvar{d}\varepsilon-\frac{1}{8\pi G}\int_{\partial\mathcal{N}}\varepsilon\,\bbvar{d}\lambda+\\
&\nonumber\quad+\frac{1}{8\pi G}\int_{\mathcal{N}}(\mathcal{L}_{\mathring{\ell}}\varepsilon)\wedge\bbvar{d}({\di\mathcal{U}})+
\frac{1}{8\pi\gamma G}\int_{\mathcal{N}}\varepsilon\wedge\bbvar{d}(\di\mathcal{U}\mathring{\varphi})+\\
&\quad+\frac{1}{8\pi\gamma G}\int_{\mathcal{N}}\left[\frac{1}{2}{\vartheta}_\smallcirc\,\di\mathcal{U}\wedge m\wedge\bbvar{d}\bar{m}-\I(\gamma+\I){\sigma}_\smallcirc\,\di\mathcal{U}\wedge\bar{m}\wedge\bbvar{d}\bar{m}+\CC\right],\label{app:theta-N3}
\end{align}
where $\varepsilon=-\I m\wedge\bar{m}$ is the area two-form on $\mathcal{N}$. We rearrange terms and obtain 
\begin{equation}
\Theta^{\mtext{Holst}}_{\mathcal{N}}=\theta_{\mathcal{C}_+}-\theta_{\mathcal{C}_-}+\frac{1}{8\pi G}\int_{\mathcal{C}_-}d^2v_o\,\Omega^2\bbvar{d}\lambda-\frac{1}{16\pi G}\int_{\mathcal{C}_+\cup\mathcal{C}_-^{-1}}\bbvar{d}\varepsilon+{\Theta}_{\mathcal{N}}.
\end{equation}
 The first three terms are corner integrals that play no role in the analysis of this paper. They define Heisenberg charges on the light cone from which one obtains simple expressions for the charges for dilatations and superrotations \cite{Wieland:2017cmf,Wieland:2021vef,Wieland:2017zkf}. The fourth term \eref{app:theta-N3} is a total derivative on field space which cannot affect the Poisson commutation relations. For the bulk of this paper, only the last term matters. 
In terms of the $SU(1,1)$ variables that we introduced in \hyperref[sec2]{Section 2} above, we obtain
\begin{align}\nonumber
\Theta_{\mathcal{N}}=&-\frac{1}{8\pi G}\int_{\mathcal{C}_+}d^2v_o\,\Omega^2\bbvar{d}\lambda+\frac{1}{8\pi G}\int_{\mathcal{N}}d^2v_o\wedge\frac{\di}{\di\mathcal{U}}\Omega^2\,\bbvar{d}(\di\mathcal{U})+\\
&+\frac{1}{8\pi\gamma G}\int_{\mathcal{N}}d^2v_o\,\Omega^2\wedge\bbvar{d}\left(\di\mathcal{U}\varphi_\smallcirc\right)+
\int_{\mathcal{N}}\di\mathcal{U}\wedge d^2v_o\,\mathrm{Tr}\left(\Pi_\smallcirc(\bbvar{d}S)S^{-1}\right).\label{app:theta-N4}
\end{align}
In here, the trace is taken with respect to the $\mathfrak{su}(1,1)$ basis \eref{su1-1-basis}. The $\mathfrak{su}(1,1)$-momentum variable admits the decomposition
\begin{equation}
\Pi_\smallcirc= L_\smallcirc J+c_\smallcirc\bar{X}+\bar{c}_\smallcirc X,
\end{equation}
where
\begin{align}
L_\smallcirc&=-\frac{1}{16\pi\gamma G}\frac{\di}{\di\mathcal{U}}\Omega^2,\label{app:Lcons}\\
c_\smallcirc&=-\frac{1}{8\pi\gamma G}(\gamma+\I)\Omega^2\sigma_\smallcirc.\label{app:ccons}
\end{align}
The last two equations are to be imposed as constraints on the phase space defined by the symplectic potential \eref{app:theta-N4}. The only other further condition on this phase space is the Raychaudhuri constraint \eref{Raychudhri-eq2}.

\providecommand{\href}[2]{#2}\begingroup\raggedright\endgroup


\begin{thebibliography}{100}

\bibitem{Wieland:2024dop}
W.~Wieland, ``{Quantum geometry of the null cone},''
  \href{http://arXiv.org/abs/2401.17491}{{\tt arXiv:2401.17491}}.

\bibitem{Wieland:2025LP}
W.~Wieland, ``Evidence for Planck luminosity bound in quantum gravity,'' {\em
  {Class. Quant. Grav.}} {\bf 42} (2025), no.~6, 06LT01.

\bibitem{Aichelburg1971}
P.~C. Aichelburg and R.~U. Sexl, ``On the gravitational field of a massless
  particle,'' {\em General Relativity and Gravitation} {\bf 2} (1971), no.~4,
  303--312.

\bibitem{Balasin:2007gh}
H.~Balasin and P.~C. Aichelburg, ``{Canonical formulation of pp-waves},'' {\em
  Gen. Rel. Grav.} {\bf 39} (2007) 1075--1085,
  \href{http://arXiv.org/abs/0705.0228}{{\tt arXiv:0705.0228}}.

\bibitem{Luk:2012hi}
J.~Luk and I.~Rodnianski, ``{Local Propagation of Impulsive
  GravitationalWaves},'' {\em Commun. Pure Appl. Math.} {\bf 68} (2015)
  511--624, \href{http://arXiv.org/abs/1209.1130}{{\tt arXiv:1209.1130}}.

\bibitem{Griffiths:1991zp}
J.~B. Griffiths, {\em {Colliding plane waves in general relativity}}.
\newblock Oxford University Press,
1991.
\newblock

\bibitem{GRColombBook}
R.~Steinbauer, M.~Grosser, M.~Kunzinger, and M.~Oberguggenberger, {\em
  Geometric Theory of Generalized Functions: with Applications to General
  Relativity}.
\newblock Mathematics and its applications. Springer, 2001.

\bibitem{Immirziparam}
G.~Immirzi, ``{Real and complex connections for canonical gravity},'' {\em
  Class. Quant. Grav.} {\bf 14} (1997) L177--L181,
\href{http://arXiv.org/abs/gr-qc/9612030}{{\tt arXiv:gr-qc/9612030}}.

\bibitem{Barbero1994}
J.~F. Barbero~G., ``{Real Ashtekar variables for Lorentzian signature space
  times},'' {\em Phys. Rev. D} {\bf 51} (1995) 5507--5510,
\href{http://arXiv.org/abs/gr-qc/9410014}{{\tt arXiv:gr-qc/9410014}}.

\bibitem{Ashtekar:2021kfp}
A.~Ashtekar and E.~Bianchi, ``{A short review of loop quantum gravity},'' {\em
  Rept. Prog. Phys.} {\bf 84} (2021), no.~4, 042001,
  \href{http://arXiv.org/abs/2104.04394}{{\tt arXiv:2104.04394}}.

\bibitem{status}
A.~Ashtekar and J.~Lewandowski, ``{Background independent quantum gravity: a
  status report},'' {\em Class. Quant. Grav.} {\bf 21} (2004), no.~15,
  R53--R152, \href{http://arXiv.org/abs/gr-qc/0404018v2}{{\tt
  arXiv:gr-qc/0404018v2}}.

\bibitem{thiemann}
C.~Thiemann, {\em Introduction to Modern Canonical Quantum General Relativity}.
\newblock Cambridge University Press, 2007.

\bibitem{rovelli}
C.~Rovelli, {\em Quantum Gravity}.
\newblock Cambridge University Press, Cambridge, 2008.

\bibitem{Ashtekar:2000eq}
A.~Ashtekar, J.~C. Baez, and K.~Krasnov, ``{Quantum geometry of isolated
  horizons and black hole entropy},'' {\em Adv. Theor. Math. Phys.} {\bf 4}
  (2000) 1--94,
\href{http://arXiv.org/abs/gr-qc/0005126}{{\tt arXiv:gr-qc/0005126}}.

\bibitem{isohorizon}
A.~Ashtekar and B.~Krishnan, ``{Isolated and Dynamical Horizons and Their
  Applications},'' {\em Living Reviews in Relativity} {\bf 7} (2004), no.~10,
  \href{http://arXiv.org/abs/gr-qc/0407042}{{\tt arXiv:gr-qc/0407042}}.

\bibitem{Ashtekar:1999wa}
A.~Ashtekar, A.~Corichi, and K.~Krasnov, ``{Isolated horizons: The Classical
  phase space},'' {\em Adv. Theor. Math. Phys.} {\bf 3} (1999) 419--478,
  \href{http://arXiv.org/abs/gr-qc/9905089}{{\tt arXiv:gr-qc/9905089}}.

\bibitem{Ashtekar:aa}
A.~Ashtekar, C.~Beetle, and S.~Fairhurst, ``Isolated Horizons: A Generalization
  of Black Hole Mechanics,'' {\em Class. Quant. Grav.} {\bf 16} (1999),
  no.~L1--L7, \href{http://arXiv.org/abs/gr-qc/9812065}{{\tt
  arXiv:gr-qc/9812065}}.

\bibitem{BarberoG.:2012ae}
J.~F. Barbero~G., J.~Lewandowski, and E.~J.~S. Villasenor, ``{Quantum isolated
  horizons and black hole entropy},'' {\em PoS} {\bf QGQGS2011} (2011) 023,
\href{http://arXiv.org/abs/1203.0174}{{\tt arXiv:1203.0174}}.

\bibitem{DiazPoloPranzetti}
J.~Diaz-Polo and D.~Pranzetti, ``{Isolated Horizons and Black Hole Entropy In
  Loop Quantum Gravity},'' {\em SIGMA} {\bf 8} (2012) 048,
\href{http://arXiv.org/abs/1112.0291}{{\tt arXiv:1112.0291}}.

\bibitem{Ashtekar:2001is}
A.~Ashtekar, C.~Beetle, and J.~Lewandowski, ``{Mechanics of rotating isolated
  horizons},'' {\em Phys. Rev. D} {\bf 64} (2001) 044016,
\href{http://arXiv.org/abs/gr-qc/0103026}{{\tt arXiv:gr-qc/0103026}}.

\bibitem{Engle:2010kt}
J.~Engle, K.~Noui, A.~Perez, and D.~Pranzetti, ``{Black hole entropy from an
  SU(2)-invariant formulation of Type I isolated horizons},'' {\em Phys. Rev.}
  {\bf D82} (2010) 044050,
\href{http://arXiv.org/abs/1006.0634}{{\tt arXiv:1006.0634}}.

\bibitem{Girelli:2005ii}
F.~Girelli and E.~R. Livine, ``{Reconstructing quantum geometry from quantum
  information: Spin networks as harmonic oscillators},'' {\em Class. Quant.
  Grav.} {\bf 22} (2005) 3295--3314,
\href{http://arXiv.org/abs/gr-qc/0501075}{{\tt arXiv:gr-qc/0501075}}.

\bibitem{Bianchi:2016hmk}
E.~Bianchi, J.~Guglielmon, L.~Hackl, and N.~Yokomizo, ``{Loop expansion and the
  bosonic representation of loop quantum gravity},'' {\em Phys. Rev. D} {\bf
  94} (2016) 086009,
\href{http://arXiv.org/abs/1609.02219}{{\tt arXiv:1609.02219}}.

\bibitem{Borja:2010rc}
E.~F. Borja, L.~Freidel, I.~Garay, and E.~R. Livine, ``{U(N) tools for Loop
  Quantum Gravity: The Return of the Spinor},'' {\em Class. Quant. Grav.} {\bf
  28} (2011) 055005,
\href{http://arXiv.org/abs/1010.5451}{{\tt arXiv:1010.5451}}.

\bibitem{twist}
L.~Freidel and S.~Speziale, ``{Twistors to twisted geometries},'' {\em Phys.
  Rev. D} {\bf 82} (2010), no.~8, 084041,
  \href{http://arXiv.org/abs/1006.0199}{{\tt arXiv:1006.0199}}.

\bibitem{twistconslor}
M.~Dupuis, L.~Freidel, E.~R. Livine, and S.~Speziale, ``{Holomorphic Lorentzian
  Simplicity Constraints},'' {\em J. Math. Phys.} {\bf 53} (2012) 032502,
  \href{http://arXiv.org/abs/1107.5274}{{\tt arXiv:1107.5274}}.

\bibitem{komplexspinors}
W.~Wieland, ``{Twistorial phase space for complex Ashtekar variables},'' {\em
  {Class. Quant. Grav.}} {\bf 29} (2011) 045007,
  \href{http://arXiv.org/abs/1107.5002}{{\tt arXiv:1107.5002}}.

\bibitem{Balachandran:1995qa}
A.~P. Balachandran, L.~Chandar, and A.~Momen, ``{Edge states in canonical
  gravity},'' in {\em {17th Annual MRST (Montreal-Rochester-Syracuse-Toronto)
  Meeting on High-energy Physics Rochester, New York, May 8-9, 1995}}.
\newblock 1995.
\newblock
\href{http://arXiv.org/abs/gr-qc/9506006}{{\tt arXiv:gr-qc/9506006}}.
\newblock

\bibitem{PhysRevD.51.632}
S.~Carlip, ``Statistical mechanics of the (2+1)-dimensional black hole,'' {\em
  Phys. Rev. D} {\bf 51} (Jan, 1995) 632--637.

\bibitem{Donnelly:2016auv}
W.~Donnelly and L.~Freidel, ``{Local subsystems in gauge theory and gravity},''
  {\em JHEP} {\bf 09} (2016) 102, \href{http://arXiv.org/abs/1601.04744}{{\tt
  arXiv:1601.04744}}.

\bibitem{Gomes:2016mwl}
H.~Gomes and A.~Riello, ``{The observer's ghost: notes on a field space
  connection},'' {\em JHEP} {\bf 05} (2017) 017,
\href{http://arXiv.org/abs/1608.08226}{{\tt arXiv:1608.08226}}.

\bibitem{Geiller:2017xad}
M.~Geiller, ``{Edge modes and corner ambiguities in 3d Chern-Simons theory and
  gravity},'' {\em Nucl. Phys. B} {\bf 924} (2017) 312--365,
\href{http://arXiv.org/abs/1703.04748}{{\tt arXiv:1703.04748}}.

\bibitem{Speranza:2017gxd}
A.~J. Speranza, ``{Local phase space and edge modes for
  diffeomorphism-invariant theories},'' {\em JHEP} {\bf 02} (2018) 021,
  \href{http://arXiv.org/abs/1706.05061}{{\tt arXiv:1706.05061}}.

\bibitem{Geiller:2017whh}
M.~Geiller, ``{Lorentz-diffeomorphism edge modes in 3d gravity},'' {\em JHEP}
  {\bf 02} (2018) 029, \href{http://arXiv.org/abs/1712.05269}{{\tt
  arXiv:1712.05269}}.

\bibitem{Wieland:2017cmf}
W.~Wieland, ``{Fock representation of gravitational boundary modes and the
  discreteness of the area spectrum},'' {\em Ann. Henri Poincar{\'e}} {\bf 18}
  (2017) 3695--3717,
\href{http://arXiv.org/abs/1706.00479}{{\tt arXiv:1706.00479}}.

\bibitem{Wieland:2017zkf}
W.~Wieland, ``{New boundary variables for classical and quantum gravity on a
  null surface},'' {\em Class. Quantum Grav.} {\bf 34} (2017) 215008,
\href{http://arXiv.org/abs/1704.07391}{{\tt arXiv:1704.07391}}.

\bibitem{Donnelly:2016rvo}
W.~Donnelly and S.~B. Giddings, ``{Observables, gravitational dressing, and
  obstructions to locality and subsystems},'' {\em Phys. Rev. D} {\bf 94}
  (2016), no.~10, 104038,
\href{http://arXiv.org/abs/1607.01025}{{\tt arXiv:1607.01025}}.

\bibitem{Takayanagi:2019tvn}
T.~Takayanagi and K.~Tamaoka, ``{Gravity Edges Modes and Hayward Term},'' {\em
  JHEP} {\bf 02} (2020) 167, \href{http://arXiv.org/abs/1912.01636}{{\tt
  arXiv:1912.01636}}.

\bibitem{Freidel:2019ees}
L.~Freidel, E.~R. Livine, and D.~Pranzetti, ``{Gravitational edge modes: from
  Kac--Moody charges to Poincar{\'e} networks},'' {\em Class. Quant. Grav.}
  {\bf 36} (2019), no.~19, 195014,
\href{http://arXiv.org/abs/1906.07876}{{\tt arXiv:1906.07876}}.

\bibitem{Francois:2021aa}
J.~Fran{\c c}ois, ``Bundle geometry of the connection space, covariant
  Hamiltonian formalism, the problem of boundaries in gauge theories, and the
  dressing field method,'' {\em Journal of High Energy Physics} {\bf 2021}
  (2021), no.~3, 225.

\bibitem{Freidel:2020ayo}
L.~Freidel, M.~Geiller, and D.~Pranzetti, ``{Edge modes of gravity. Part III.
  Corner simplicity constraints},'' {\em JHEP} {\bf 01} (2021) 100,
  \href{http://arXiv.org/abs/2007.12635}{{\tt arXiv:2007.12635}}.

\bibitem{Freidel:2020svx}
L.~Freidel, M.~Geiller, and D.~Pranzetti, ``{Edge modes of gravity. Part II.
  Corner metric and Lorentz charges},'' {\em JHEP} {\bf 11} (2020) 027,
  \href{http://arXiv.org/abs/2007.03563}{{\tt arXiv:2007.03563}}.

\bibitem{Freidel:2020xyx}
L.~Freidel, M.~Geiller, and D.~Pranzetti, ``{Edge modes of gravity. Part I.
  Corner potentials and charges},'' {\em JHEP} {\bf 11} (2020) 026,
  \href{http://arXiv.org/abs/2006.12527}{{\tt arXiv:2006.12527}}.

\bibitem{Donnelly:2020xgu}
W.~Donnelly, L.~Freidel, S.~F. Moosavian, and A.~J. Speranza, ``{Gravitational
  edge modes, coadjoint orbits, and hydrodynamics},'' {\em JHEP} {\bf 09}
  (2021) 008, \href{http://arXiv.org/abs/2012.10367}{{\tt arXiv:2012.10367}}.

\bibitem{Freidel:2021cjp}
L.~Freidel, R.~Oliveri, D.~Pranzetti, and S.~Speziale, ``{Extended corner
  symmetry, charge bracket and Einstein\textquoteright{}s equations},'' {\em
  JHEP} {\bf 09} (2021) 083, \href{http://arXiv.org/abs/2104.12881}{{\tt
  arXiv:2104.12881}}.

\bibitem{Carrozza:2021gju}
S.~Carrozza and P.~A. Hoehn, ``{Edge modes as reference frames and boundary
  actions from post-selection},'' {\em JHEP} {\bf 02} (2022) 172,
  \href{http://arXiv.org/abs/2109.06184}{{\tt arXiv:2109.06184}}.

\bibitem{Carrozza:2022xut}
S.~Carrozza, S.~Eccles, and P.~A. Hoehn, ``{Edge modes as dynamical frames:
  charges from post-selection in generally covariant theories},''
  \href{http://arXiv.org/abs/2205.00913}{{\tt arXiv:2205.00913}}.

\bibitem{Kabel23}
V.~Kabel, {\v C}.~Brukner, and W.~Wieland, ``Quantum reference frames at the
  boundary of spacetime,'' {\em Phys. Rev. D} {\bf 108} (Nov, 2023) 106022,
  \href{http://arXiv.org/abs/2302.11629}{{\tt arXiv:2302.11629}}.

\bibitem{Giesel:2024xtb}
K.~Giesel, V.~Kabel, and W.~Wieland, ``{Linking Edge Modes and Geometrical
  Clocks in Linearized Gravity},'' \href{http://arXiv.org/abs/2410.17339}{{\tt
  arXiv:2410.17339}}.

\bibitem{Ciambelli:2022vot}
L.~Ciambelli, ``{From Asymptotic Symmetries to the Corner Proposal},'' {\em
  PoS} {\bf Modave2022} (2023) 002, \href{http://arXiv.org/abs/2212.13644}{{\tt
  arXiv:2212.13644}}.

\bibitem{Wieland:2021vef}
W.~Wieland, ``{Gravitational SL(2, \ensuremath{\mathbb{R}}) algebra on the
  light cone},'' {\em JHEP} {\bf 07} (2021) 057,
  \href{http://arXiv.org/abs/2104.05803}{{\tt arXiv:2104.05803}}.

\bibitem{Araujo-Regado:2024dpr}
G.~Araujo-Regado, P.~A. Hoehn, F.~Sartini, and B.~Tomova, ``{Soft edges: the
  many links between soft and edge modes},''
  \href{http://arXiv.org/abs/2412.14548}{{\tt arXiv:2412.14548}}.

\bibitem{Donnay:2019jiz}
L.~Donnay and C.~Marteau, ``{Carrollian Physics at the Black Hole Horizon},''
  {\em Class. Quant. Grav.} {\bf 36} (2019), no.~16, 165002,
  \href{http://arXiv.org/abs/1903.09654}{{\tt arXiv:1903.09654}}.

\bibitem{Ciambelli:2019lap}
L.~Ciambelli, R.~G. Leigh, C.~Marteau, and P.~M. Petropoulos, ``{Carroll
  Structures, Null Geometry and Conformal Isometries},'' {\em Phys. Rev. D}
  {\bf 100} (2019), no.~4, 046010, \href{http://arXiv.org/abs/1905.02221}{{\tt
  arXiv:1905.02221}}.

\bibitem{Ciambelli:2018ojf}
L.~Ciambelli and C.~Marteau, ``{Carrollian conservation laws and Ricci-flat
  gravity},'' {\em Class. Quant. Grav.} {\bf 36} (2019), no.~8, 085004,
  \href{http://arXiv.org/abs/1810.11037}{{\tt arXiv:1810.11037}}.

\bibitem{Lehner:2016vdi}
L.~Lehner, R.~C. Myers, E.~Poisson, and R.~D. Sorkin, ``{Gravitational action
  with null boundaries},'' {\em Phys. Rev. D} {\bf 94} (2016), no.~8, 084046,
  \href{http://arXiv.org/abs/1609.00207}{{\tt arXiv:1609.00207}}.

\bibitem{Parattu:2015gga}
K.~Parattu, S.~Chakraborty, B.~R. Majhi, and T.~Padmanabhan, ``{A Boundary Term
  for the Gravitational Action with Null Boundaries},'' {\em Gen. Rel. Grav.}
  {\bf 48} (2016), no.~7, 94, \href{http://arXiv.org/abs/1501.01053}{{\tt
  arXiv:1501.01053}}.

\bibitem{Hopfmuller:2016scf}
F.~Hopfm{\"u}ller and L.~Freidel, ``{Gravity Degrees of Freedom on a Null
  Surface},'' {\em Phys. Rev.} {\bf D95} (2017), no.~10, 104006,
\href{http://arXiv.org/abs/1611.03096}{{\tt arXiv:1611.03096}}.

\bibitem{Freidel:2022vjq}
L.~Freidel and P.~Jai-akson, ``{Carrollian hydrodynamics and symplectic
  structure on stretched horizons},''
  \href{http://arXiv.org/abs/2211.06415}{{\tt arXiv:2211.06415}}.

\bibitem{Adami:2021nnf}
H.~Adami, D.~Grumiller, M.~M. Sheikh-Jabbari, V.~Taghiloo, H.~Yavartanoo, and
  C.~Zwikel, ``{Null boundary phase space: slicings, news \& memory},'' {\em
  JHEP} {\bf 11} (2021) 155, \href{http://arXiv.org/abs/2110.04218}{{\tt
  arXiv:2110.04218}}.

\bibitem{Ciambelli:2023mir}
L.~Ciambelli, L.~Freidel, and R.~G. Leigh, ``{Null Raychaudhuri: Canonical
  Structure and the Dressing Time},''
  \href{http://arXiv.org/abs/2309.03932}{{\tt arXiv:2309.03932}}.

\bibitem{Reisenberger:2018xkn}
M.~P. Reisenberger, ``{The Poisson brackets of free null initial data for
  vacuum general relativity},'' {\em Class. Quant. Grav.} {\bf 35} (2018),
  no.~18, 185012,
\href{http://arXiv.org/abs/1804.10284}{{\tt arXiv:1804.10284}}.

\bibitem{Fuchs:2017jyk}
A.~Fuchs and M.~P. Reisenberger, ``{Integrable structures and the quantization
  of free null initial data for gravity},'' {\em Class. Quant. Grav.} {\bf 34}
  (2017), no.~18, 185003, \href{http://arXiv.org/abs/1704.06992}{{\tt
  arXiv:1704.06992}}.

\bibitem{Reisenberger:2012zq}
M.~P. Reisenberger, ``{The symplectic 2-form for gravity in terms of free null
  initial data},'' {\em Class. Quant. Grav.} {\bf 30} (2013) 155022,
  \href{http://arXiv.org/abs/1211.3880}{{\tt arXiv:1211.3880}}.

\bibitem{AndradeeSilva:2022iic}
R.~Andrade~e Silva and T.~Jacobson, ``{Causal diamonds in (2+1)-dimensional
  quantum gravity},'' {\em Phys. Rev. D} {\bf 107} (2023), no.~2, 024033,
  \href{http://arXiv.org/abs/2203.10084}{{\tt arXiv:2203.10084}}.

\bibitem{Ciambelli:2024swv}
L.~Ciambelli, L.~Freidel, and R.~G. Leigh, ``{Quantum null geometry and
  gravity},'' {\em JHEP} {\bf 12} (2024) 028,
  \href{http://arXiv.org/abs/2407.11132}{{\tt arXiv:2407.11132}}.

\bibitem{Strominger:1997eq}
A.~Strominger, ``{Black hole entropy from near horizon microstates},'' {\em
  JHEP} {\bf 02} (1998) 009,
\href{http://arXiv.org/abs/hep-th/9712251}{{\tt arXiv:hep-th/9712251}}.

\bibitem{Carlip:1998wz}
S.~Carlip, ``{Black hole entropy from conformal field theory in any
  dimension},'' {\em Phys. Rev. Lett.} {\bf 82} (1999) 2828--2831,
  \href{http://arXiv.org/abs/hep-th/9812013}{{\tt arXiv:hep-th/9812013}}.

\bibitem{Carlip:1999cy}
S.~Carlip, ``{Entropy from conformal field theory at Killing horizons},'' {\em
  Class. Quant. Grav.} {\bf 16} (1999) 3327--3348,
\href{http://arXiv.org/abs/gr-qc/9906126}{{\tt arXiv:gr-qc/9906126}}.

\bibitem{Wieland:2020gno}
W.~Wieland, ``{Null infinity as an open Hamiltonian system},'' {\em JHEP} {\bf
  04} (2021) 095, \href{http://arXiv.org/abs/2012.01889}{{\tt
  arXiv:2012.01889}}.

\bibitem{Ashtekar:1981sf}
A.~Ashtekar, ``{Asymptotic Quantization of the Gravitational Field},'' {\em
  Phys. Rev. Lett.} {\bf 46} (1981)
573--576.

\bibitem{Strominger:2017zoo}
A.~Strominger, {\em {Lectures on the Infrared Structure of Gravity and Gauge
  Theory}}.
\newblock Princeton University Press, Princeton, 2018.
\newblock
\href{http://arXiv.org/abs/1703.05448}{{\tt arXiv:1703.05448}}.
\newblock

\bibitem{AshtekarNullInfinity}
A.~Ashtekar, {\em {Asymptotic Quantization}}.
\newblock Bibliopolis, Napoli, 1987.
\newblock Based on 1984 Naples Lectures.

\bibitem{Balachandran:1994up}
A.~P. Balachandran, L.~Chandar, and A.~Momen, ``{Edge states in gravity and
  black hole physics},'' {\em Nucl. Phys. B} {\bf 461} (1996) 581--596,
\href{http://arXiv.org/abs/gr-qc/9412019}{{\tt arXiv:gr-qc/9412019}}.

\bibitem{Donnelly:2015hxa}
W.~Donnelly and A.~C. Wall, ``{Geometric entropy and edge modes of the
  electromagnetic field},'' {\em Phys. Rev. D} {\bf 94} (2016), no.~10, 104053,
  \href{http://arXiv.org/abs/1506.05792}{{\tt arXiv:1506.05792}}.

\bibitem{DonnellyGiddings2016}
W.~Donnelly and S.~B. Giddings, ``{Observables, gravitational dressing, and
  obstructions to locality and subsystems},''
\href{http://arXiv.org/abs/1607.01025}{{\tt arXiv:1607.01025}}.

\bibitem{oriti}
D.~Oriti, ``{The group field theory approach to quantum gravity},'' in {\em
  Approaches to Quantum Gravity}.
\newblock Cambridge University Press, Cambridge, 2009.

\bibitem{Freidel:2005qe}
L.~Freidel, ``{Group field theory: An Overview},'' {\em Int. J. Theor. Phys.}
  {\bf 44} (2005) 1769--1783, \href{http://arXiv.org/abs/hep-th/0505016}{{\tt
  arXiv:hep-th/0505016}}.

\bibitem{Oriti:2013aqa}
D.~Oriti, ``{Group field theory as the 2nd quantization of Loop Quantum
  Gravity},'' {\em Class. Quant. Grav.} {\bf 33} (2016), no.~8, 085005,
  \href{http://arXiv.org/abs/1310.7786}{{\tt arXiv:1310.7786}}.

\bibitem{Gielen:2024sxs}
S.~Gielen, ``{Hilbert space formalisms for group field theory},''
  \href{http://arXiv.org/abs/2412.07847}{{\tt arXiv:2412.07847}}.

\bibitem{Rovelli:2014ssa}
C.~Rovelli and F.~Vidotto, {\em {Covariant Loop Quantum Gravity}}.
\newblock Cambridge University Press,
2014.
\newblock

\bibitem{alexreview}
A.~Perez, ``{The Spin-Foam Approach to Quantum Gravity},'' {\em Living Rev.
  Rel.} {\bf 16} (2013), no.~3,
\href{http://arXiv.org/abs/1205.2019}{{\tt arXiv:1205.2019}}.

\bibitem{Engle2023}
J.~Engle and S.~Speziale, ``Spin Foams: Foundations,'' in {\em Handbook of
  Quantum Gravity}, C.~Bambi, L.~Modesto, and I.~Shapiro, eds., pp.~1--40.
\newblock Springer Nature Singapore, Singapore, 2023.

\bibitem{ashtekar}
A.~Ashtekar, {\em {Lectures on Non-Pertubative Canonical Gravity}}.
\newblock World Scientific, 1991.

\bibitem{Thiemann2023}
T.~Thiemann and K.~Giesel, {\em Hamiltonian Theory: Dynamics}, pp.~1--52.
\newblock Springer Nature Singapore, Singapore, 2023.

\bibitem{Asante2023}
S.~K. Asante, B.~Dittrich, and S.~Steinhaus, {\em Spin Foams, Refinement Limit,
  and Renormalization}, pp.~1--37.
\newblock Springer Nature Singapore, Singapore, 2023.

\bibitem{Oriti:2013jga}
D.~Oriti, ``{Disappearance and emergence of space and time in quantum
  gravity},'' {\em Stud. Hist. Phil. Sci. B} {\bf 46} (2014) 186--199,
  \href{http://arXiv.org/abs/1302.2849}{{\tt arXiv:1302.2849}}.

\bibitem{FernandoBarbero:2009ai}
G.~J. Fernando~Barbero, J.~Lewandowski, and E.~J.~S. Villasenor, ``{Flux-area
  operator and black hole entropy},'' {\em Phys. Rev. D} {\bf 80} (2009)
  044016, \href{http://arXiv.org/abs/0905.3465}{{\tt arXiv:0905.3465}}.

\bibitem{Surya:2019ndm}
S.~Surya, ``{The causal set approach to quantum gravity},'' {\em Living Rev.
  Rel.} {\bf 22} (2019), no.~1, 5, \href{http://arXiv.org/abs/1903.11544}{{\tt
  arXiv:1903.11544}}.

\bibitem{shapebook}
F.~Mercati, ``Shape Dynamics: Relativity and Relationalism,''
\newblock Oxford University Press, Oxford, 2018.

\bibitem{Gomes:2010fh}
H.~Gomes, S.~Gryb, and T.~Koslowski, ``{Einstein gravity as a 3D conformally
  invariant theory},'' {\em Class. Quant. Grav.} {\bf 28} (2011) 045005,
  \href{http://arXiv.org/abs/1010.2481}{{\tt arXiv:1010.2481}}.

\bibitem{Ellis1984}
G.~F.~R. Ellis, {\em Relativistic Cosmology: Its Nature, Aims and Problems},
  pp.~215--288.
\newblock Springer Netherlands, Dordrecht, 1984.

\bibitem{penroserindler}
R.~Penrose and W.~Rindler, {\em Spinors and Space-Time, Two-Spinor Calculus and
  Relativistic Fields}, vol.~1 and 2.
\newblock Cambridge University Press, Cambridge, 1984.

\bibitem{Giesel:2012}
K.~Giesel and T.~Thiemann, ``{Scalar Material Reference Systems and Loop
  Quantum Gravity},'' {\em Class. Quant. Grav.} {\bf 32} (2015) 135015,
  \href{http://arXiv.org/abs/1206.3807}{{\tt arXiv:1206.3807}}.

\bibitem{Giesel:2007wn}
K.~Giesel and T.~Thiemann, ``{Algebraic quantum gravity (AQG). IV. Reduced
  phase space quantisation of loop quantum gravity},'' {\em Class.Quant.Grav.}
  {\bf 27} (2010) 175009,
\href{http://arXiv.org/abs/0711.0119}{{\tt arXiv:0711.0119}}.

\bibitem{Husain:2011tk}
V.~Husain and T.~Pawlowski, ``{Time and a physical Hamiltonian for quantum
  gravity},'' {\em Phys. Rev. Lett.} {\bf 108} (2012) 141301,
  \href{http://arXiv.org/abs/1108.1145}{{\tt arXiv:1108.1145}}.

\bibitem{Giesel:2016gxq}
K.~Giesel and A.~Vetter, ``{Reduced loop quantization with four
  Klein\textendash{}Gordon scalar fields as reference matter},'' {\em Class.
  Quant. Grav.} {\bf 36} (2019), no.~14, 145002,
  \href{http://arXiv.org/abs/1610.07422}{{\tt arXiv:1610.07422}}.

\bibitem{partobs}
C.~Rovelli, ``{Partial observables},'' {\em Phys. Rev. D} {\bf 65} (June, 2002)
  124013, \href{http://arXiv.org/abs/gr-qc/0110035v3}{{\tt
  arXiv:gr-qc/0110035v3}}.

\bibitem{Domagala:2010bm}
M.~Domagala, K.~Giesel, W.~Kaminski, and J.~Lewandowski, ``{Gravity quantized:
  Loop Quantum Gravity with a Scalar Field},'' {\em Phys. Rev. D} {\bf 82}
  (2010) 104038, \href{http://arXiv.org/abs/1009.2445}{{\tt arXiv:1009.2445}}.

\bibitem{spezialetwist2}
M.~Dupuis, L.~Freidel, E.~R. Livine, and S.~Speziale, ``{Holomorphic Lorentzian
  Simplicity Constraints},'' {\em J. Math. Phys.} {\bf 53} (2012) 032502,
  \href{http://arXiv.org/abs/1107.5274}{{\tt arXiv:1107.5274}}.

\bibitem{LQGvertexfinite}
J.~Engle, E.~Livine, and C.~Rovelli, ``{LQG vertex with finite Immirzi
  parameter},'' {\em Nucl. Phys. B} {\bf 799} (2008) 136--149,
  \href{http://arXiv.org/abs/0711.0146}{{\tt arXiv:0711.0146}}.

\bibitem{PhysRevD.82.064026}
B.~Dittrich and J.~P. Ryan, ``Simplicity in simplicial phase space,'' {\em
  Phys. Rev. D} {\bf 82} (Sep, 2010) 064026,
  \href{http://arXiv.org/abs/1006.4295}{{\tt arXiv:1006.4295}}.

\bibitem{flppdspinfoam}
J.~Engle, R.~Pereira, and C.~Rovelli, ``Flipped spinfoam vertex and loop
  gravity,'' {\em Nucl. Phys. B} {\bf 798} (2008) 251--290,
  \href{http://arXiv.org/abs/0708.1236v1}{{\tt arXiv:0708.1236v1}}.

\bibitem{Wieland:2013cr}
W.~Wieland, ``{Hamiltonian spinfoam gravity},'' {\em Class. Quant. Grav.} {\bf
  31} (2014) 025002,
\href{http://arXiv.org/abs/1301.5859}{{\tt arXiv:1301.5859}}.

\bibitem{Vytheeswaran:1994np}
A.~S. Vytheeswaran, ``{Gauge unfixing in second class constrained systems},''
  {\em Annals Phys.} {\bf 236} (1994)
297--324.

\bibitem{LOSTtheorem}
J.~Lewandowski, A.~Okolow, H.~Sahlmann, and T.~Thiemann, ``{Uniqueness of
  diffeomorphism invariant states on holonomy-flux algebras},'' {\em Commun.
  Math. Phys.} {\bf 267} (2006) 703--733,
\href{http://arXiv.org/abs/gr-qc/0504147}{{\tt arXiv:gr-qc/0504147}}.

\bibitem{Ashtekar:1994mh}
A.~Ashtekar and J.~Lewandowski, ``{Projective techniques and functional
  integration for gauge theories},'' {\em J. Math. Phys.} {\bf 36} (1995)
  2170--2191,
\href{http://arXiv.org/abs/gr-qc/9411046}{{\tt arXiv:gr-qc/9411046}}.

\bibitem{Thiemann:1997rq}
T.~Thiemann, ``{Kinematical Hilbert spaces for Fermionic and Higgs quantum
  field theories},'' {\em Class. Quant. Grav.} {\bf 15} (1998) 1487--1512,
  \href{http://arXiv.org/abs/gr-qc/9705021}{{\tt arXiv:gr-qc/9705021}}.

\bibitem{polchinski}
J.~Polchinski, {\em String Theory}, vol.~1 and 2.
\newblock Cambridge University Press, 2005.

\bibitem{green}
M.~Green, J.~Schwarz, and E.~Witten, {\em Superstring Theory}, vol.~1 and 2.
\newblock Cambridge University Press, 1988.

\bibitem{Misner:1973prb}
C.~W. Misner, K.~S. Thorne, and J.~A. Wheeler, {\em {Gravitation}}.
\newblock W. H. Freeman, San Francisco, 1973.

\bibitem{Giulini:2009np}
D.~Giulini, ``{The Superspace of Geometrodynamics},'' {\em Gen. Rel. Grav.}
  {\bf 41} (2009) 785--815, \href{http://arXiv.org/abs/0902.3923}{{\tt
  arXiv:0902.3923}}.

\bibitem{DeWitt:1967yk}
B.~S. DeWitt, ``{Quantum Theory of Gravity. 1},'' {\em Phys. Rev.} {\bf 160}
  (1967) 1113--1148.

\bibitem{Rovelli:2010qx}
C.~Rovelli and M.~Smerlak, ``{In quantum gravity, summing is refining},'' {\em
  Class. Quant. Grav.} {\bf 29} (2012) 055004,
\href{http://arXiv.org/abs/1010.5437}{{\tt arXiv:1010.5437}}.

\bibitem{Bahr:2012qj}
B.~Bahr, B.~Dittrich, F.~Hellmann, and W.~Kaminski, ``{Holonomy Spin Foam
  Models: Definition and Coarse Graining},'' {\em Phys. Rev. D} {\bf 87} (2013)
  044048,
\href{http://arXiv.org/abs/1208.3388}{{\tt arXiv:1208.3388}}.

\bibitem{Dittrich:2014ala}
B.~Dittrich, ``{The continuum limit of loop quantum gravity - a framework for
  solving the theory},'' in {\em Loop Quantum Gravity, The First Thirty Years},
  A.~Abhay and J.~Pullin, eds., vol.~4.
\newblock World Scientific, 2017.
\newblock
\href{http://arXiv.org/abs/1409.1450}{{\tt arXiv:1409.1450}}.
\newblock

\bibitem{Cao:2016mst}
C.~Cao, S.~M. Carroll, and S.~Michalakis, ``{Space from Hilbert Space:
  Recovering Geometry from Bulk Entanglement},'' {\em Phys. Rev. D} {\bf 95}
  (2017), no.~2, 024031, \href{http://arXiv.org/abs/1606.08444}{{\tt
  arXiv:1606.08444}}.

\bibitem{CFTbook}
D.~S. Philippe Di~Francecso, Pierre~Mathieu, {\em Conformal Field Theory}.
\newblock Springer, 1997.

\bibitem{Wiltshire:2007jk}
D.~L. Wiltshire, ``{Cosmic clocks, cosmic variance and cosmic averages},'' {\em
  New J. Phys.} {\bf 9} (2007) 377,
  \href{http://arXiv.org/abs/gr-qc/0702082}{{\tt arXiv:gr-qc/0702082}}.

\bibitem{Ashtekar:2024xwl}
A.~Ashtekar and S.~Speziale, ``{The Operational Meaning of Total Energy of
  Isolated Systems in General Relativity},'' {\em Universe} {\bf 10} (2024),
  no.~9, 367, \href{http://arXiv.org/abs/2409.06698}{{\tt arXiv:2409.06698}}.

\bibitem{ADMmass}
R.~Arnowitt, S.~Deser, and C.~W. Misner, ``Coordinate Invariance and Energy
  Expressions in General Relativity,'' {\em Phys. Rev.} {\bf 122} (1961)
  997--1006.

\bibitem{Schon:1979rg}
R.~Schon and S.-T. Yau, ``{On the Proof of the positive mass conjecture in
  general relativity},'' {\em Commun. Math. Phys.} {\bf 65} (1979)
45--76.

\bibitem{Wittenproof}
E.~Witten, ``{A Simple Proof of the Positive Energy Theorem},'' {\em Commun.
  Math. Phys.} {\bf 80} (1981)
381.

\bibitem{Bondi21}
H.~Bondi, M.~G.~J. van~der Burg, and A.~W.~K. Metzner, ``Gravitational waves in
  general relativity, VII. Waves from axi-symmetric isolated system,'' {\em
  Proc. of the Royal Soc. Lond. A: Mathematical, Physical and Engineering
  Sciences} {\bf 269} (1962), no.~1336, 21--52.

\bibitem{Sachs103}
R.~Sachs, ``Gravitational waves in general relativity VIII. Waves in
  asymptotically flat space-time,'' {\em Proceedings of the Royal Society
  London A} {\bf 270} (1962), no.~1340, 103--126.

\bibitem{holst}
S.~Holst, ``{Barbero's Hamiltonian derived from a generalized Hilbert-Palatini
  action},'' {\em Phys. Rev. D} {\bf 53} (November, 1996) 5966--5969,
  \href{http://arXiv.org/abs/gr-qc/9511026}{{\tt arXiv:gr-qc/9511026}}.

\bibitem{Parviol}
R.~Hojman, C.~Mukku, and W.~A. Sayed, ``Parity violation in metric-torsion
  theories of gravitation,'' {\em Phys. Rev. D} {\bf 22} (Oct, 1980)
  1915--1921.

\bibitem{surholst}
A.~Corichi and E.~Wilson-Ewing, ``{Surface terms, asymptotics and
  thermodynamics of the Holst action},'' {\em Class. Quantum Grav.} {\bf 27}
  (2010), no.~20, \href{http://arXiv.org/abs/1005.3298}{{\tt arXiv:1005.3298}}.

\bibitem{Rovelli:1990pi}
C.~Rovelli, ``{Quantum Reference Frames},'' {\em Class. Quant. Grav.} {\bf 8}
  (1991) 317--332.

\bibitem{Giacomini:2017zju}
F.~Giacomini, E.~Castro-Ruiz, and {\v C}.~Brukner, ``{Quantum mechanics and the
  covariance of physical laws in quantum reference frames},'' {\em Nature
  Commun.} {\bf 10} (2019), no.~1, 494,
  \href{http://arXiv.org/abs/1712.07207}{{\tt arXiv:1712.07207}}.

\bibitem{Loveridge2018}
L.~Loveridge, T.~Miyadera, and P.~Busch, ``Symmetry, Reference Frames, and
  Relational Quantities in Quantum Mechanics,'' {\em Foundations of Physics}
  {\bf 48} (2018), no.~2, 135--198.

\bibitem{Vanrietvelde:2018pgb}
A.~Vanrietvelde, P.~A. Hoehn, F.~Giacomini, and E.~Castro-Ruiz, ``{A change of
  perspective: switching quantum reference frames via a perspective-neutral
  framework},'' {\em Quantum} {\bf 4} (2020) 225,
  \href{http://arXiv.org/abs/1809.00556}{{\tt arXiv:1809.00556}}.

\bibitem{Hoehn:2019owq}
P.~A. H\"ohn, A.~R. Smith, and M.~P. Lock, ``{The Trinity of Relational Quantum
  Dynamics},'' \href{http://arXiv.org/abs/1912.00033}{{\tt arXiv:1912.00033}}.

\bibitem{delaHamette:2021oex}
A.-C. de~la Hamette, T.~D. Galley, P.~A. Hoehn, L.~Loveridge, and M.~P.
  Mueller, ``{Perspective-neutral approach to quantum frame covariance for
  general symmetry groups},'' \href{http://arXiv.org/abs/2110.13824}{{\tt
  arXiv:2110.13824}}.

\bibitem{Rovelliarea}
C.~Rovelli and L.~Smolin, ``Discreteness of area and volume in quantum
  gravity,'' {\em Nuclear Physics B} {\bf 442} (1995), no.~3, 593--619,
  \href{http://arXiv.org/abs/gr-qc/9411005}{{\tt arXiv:gr-qc/9411005}}.

\bibitem{AshtekarLewandowskiArea}
A.~Ashtekar and J.~Lewandowski, ``{Quantum theory of geometry I.: Area
  operators},'' {\em Class. Quant. Grav.} {\bf 14} (1997) A55--A82,
\href{http://arXiv.org/abs/gr-qc/9602046}{{\tt arXiv:gr-qc/9602046}}.

\bibitem{Haggard:2023tnj}
H.~M. Haggard, J.~Lewandowski, and H.~Sahlmann, ``{Emergence of Riemannian
  Quantum Geometry},'' in {\em Handbook of Quantum Gravity}, C.~Bambi,
  L.~Modesto, and I.~Shapiro, eds.
\newblock Springer, 2023.
\newblock \href{http://arXiv.org/abs/2302.02840}{{\tt arXiv:2302.02840}}.

\bibitem{bianchisommer}
E.~Bianchi and H.~M. Haggard, ``{Discreteness of the volume of space from
  Bohr-Sommerfeld quantization},'' {\em Phys. Rev. Lett.} {\bf 107} (2011)
  011301,
\href{http://arXiv.org/abs/1102.5439}{{\tt arXiv:1102.5439}}.

\bibitem{Beetle:2010rd}
C.~Beetle and J.~Engle, ``{Generic isolated horizons in loop quantum
  gravity},'' {\em Class. Quant. Grav.} {\bf 27} (2010) 235024,
  \href{http://arXiv.org/abs/1007.2768}{{\tt arXiv:1007.2768}}.

\bibitem{Krasnov:1998mp}
K.~Krasnov, ``{Note on the area spectrum in quantum gravity},'' {\em Class.
  Quant. Grav.} {\bf 15} (1998) L47--L53,
  \href{http://arXiv.org/abs/gr-qc/9803074}{{\tt arXiv:gr-qc/9803074}}.

\bibitem{Wieland:2017ksn}
W.~Wieland, ``{Quantum gravity in three dimensions, Witten spinors and the
  quantisation of length},'' {\em Nucl. Phys. B} {\bf 930} (2018) 219--234,
\href{http://arXiv.org/abs/1711.01276}{{\tt arXiv:1711.01276}}.

\bibitem{shadowstats}
A.~Ashtekar, S.~Fairhurst, and J.~L. Willis, ``{Quantum gravity, shadow states,
  and quantum mechanics},'' {\em Class. Quantum Grav.} {\bf 20} (March, 2003)
  1031--1061, \href{http://arXiv.org/abs/gr-qc/0207106v3}{{\tt
  arXiv:gr-qc/0207106v3}}.

\bibitem{Freidel:2010tt}
L.~Freidel and E.~R. Livine, ``{U(N) Coherent States for Loop Quantum
  Gravity},'' {\em J. Math. Phys.} {\bf 52} (2011) 052502,
\href{http://arXiv.org/abs/1005.2090}{{\tt arXiv:1005.2090}}.

\bibitem{Freidel:2010aq}
L.~Freidel and S.~Speziale, ``{Twisted geometries: A geometric parametrisation
  of SU(2) phase space},'' {\em Phys. Rev.} {\bf D82} (2010) 084040,
\href{http://arXiv.org/abs/1001.2748}{{\tt arXiv:1001.2748}}.

\bibitem{Sahlmann:2001nv}
H.~Sahlmann, T.~Thiemann, and O.~Winkler, ``{Coherent states for canonical
  quantum general relativity and the infinite tensor product extension},'' {\em
  Nucl. Phys. B} {\bf 606} (2001) 401--440,
  \href{http://arXiv.org/abs/gr-qc/0102038}{{\tt arXiv:gr-qc/0102038}}.

\bibitem{Thiemann:2000bw}
T.~Thiemann, ``{Gauge field theory coherent states (GCS): 1. General
  properties},'' {\em Class. Quant. Grav.} {\bf 18} (2001) 2025--2064,
  \href{http://arXiv.org/abs/hep-th/0005233}{{\tt arXiv:hep-th/0005233}}.

\bibitem{Bianchi:2012ev}
E.~Bianchi and R.~C. Myers, ``{On the Architecture of Spacetime Geometry},''
  {\em Class. Quant. Grav.} {\bf 31} (2014) 214002,
\href{http://arXiv.org/abs/1212.5183}{{\tt arXiv:1212.5183}}.

\bibitem{selfdualtwo}
R.~Capovilla, T.~Jacobson, J.~Dell, and L.~Mason, ``{Selfdual two forms and
  gravity},'' {\em Class. Quant. Grav.} {\bf 8} (1991)
41--57.

\bibitem{komplex1}
W.~Wieland, ``{Complex Ashtekar Variables and Reality Conditions for Holst's
  Action},'' {\em {Annales Henri Poincar{\'e}}} {\bf 13} (2012), no.~2,
  425--448, \href{http://arXiv.org/abs/1012.1738}{{\tt arXiv:1012.1738}}.

\bibitem{Wieland:2019hkz}
W.~Wieland, ``{Generating functional for gravitational null initial data},''
  {\em Class. Quant. Grav.} {\bf 36} (2019), no.~23, 235007,
\href{http://arXiv.org/abs/1905.06357}{{\tt arXiv:1905.06357}}.

\bibitem{Freidel:2024emv}
L.~Freidel and P.~Jai-akson, ``{Geometry of Carrollian stretched horizons},''
  {\em Class. Quant. Grav.} {\bf 42} (2025), no.~6, 065010,
  \href{http://arXiv.org/abs/2406.06709}{{\tt arXiv:2406.06709}}.

\bibitem{Gomes:2019xto}
H.~Gomes and A.~Riello, ``{The quasilocal degrees of freedom of Yang-Mills
  theory},'' {\em SciPost Phys.} {\bf 10} (2021), no.~6, 130,
  \href{http://arXiv.org/abs/1910.04222}{{\tt arXiv:1910.04222}}.

\end{thebibliography}
\end{document}